\documentclass[fleqn,usenatbib]{mnras}
\usepackage{newtxtext,newtxmath}

\usepackage[T1]{fontenc}
\usepackage{ae,aecompl}
\usepackage[normalem]{ulem}
\usepackage{graphicx}	
\usepackage{amsmath}	
\usepackage{amssymb}	
\usepackage[utf8]{inputenc}
\usepackage{lastpage}

\title[Bulge Formation in the Auriga galaxies]{The prevalence of pseudo-bulges in the Auriga simulations}

\author[I.D. Gargiulo et al.]{Ignacio D. Gargiulo$^{1,2}$\thanks{E-mail: gargiulo@dfuls.cl},
Antonela Monachesi$^{1,2}$, Facundo A. G\'omez$^{1,2}$, \newauthor{Robert J. J. Grand$^{3}$},  Federico Marinacci$^{4}$, R\"udiger Pakmor$^{3}$, \newauthor Simon D. M. White$^{3}$, Eric F. Bell$^{5}$, Francesca Fragkoudi$^{3}$,  and Patricia Tissera$^{6,7}$
\vspace{0.2cm}\\
\\
$^{1}$Instituto de Investigaci\'on Multidisciplinar en Ciencia y Tecnolog\'ia, Universidad de La Serena, Ra\'ul Bitr\'an 1305, La Serena, Chile\\
$^{2}$Departamento de F\'isica y Astronom\'ia, Universidad de La Serena, Av. Juan Cisternas 1200 Norte, La Serena, Chile\\
$^{3}$Max-Planck-Institut f\"{u}r Astrophysik, Karl-Schwarzschild-Str. 1, D-85748, Garching, Germany\\
$^{4}$Department of Physics and Astronomy, University of Bologna, via Gobetti 93/2, 40129 Bologna, Italy\\
$^{5}$Department of Astronomy, University of Michigan, 1085 S. University Avenue, Ann Arbor, MI 48109, USA\\
$^{6}$Departamento de Ciencias F\'isicas, Universidad Andr\'es Bello, Fernandez Concha 700, Santiago, Chile\\
$^{7}$Millennium Institute of Astrophysics, Fernandez Concha 700, Santiago, Chile
}

\date{Accepted XXX. Received YYY; in original form ZZZ}

\pubyear{2018}

\begin{document}
\label{firstpage}
\pagerange{\pageref{firstpage}--\pageref{lastpage}}
\maketitle

\begin{abstract}

We study the galactic bulges in the Auriga simulations, a suite of thirty cosmological magneto-hydrodynamical zoom-in simulations of late-type galaxies in Milky Way-sized dark matter haloes performed with the moving-mesh code AREPO. We aim to characterize bulge formation mechanisms in this large suite of galaxies simulated at high resolution in a fully cosmological context. The bulges of the Auriga galaxies show a large variety in their shapes, sizes and formation histories. According to observational classification criteria, such as S\'ersic index and degree of ordered rotation, the majority of the Auriga bulges can be classified as pseudo-bulges, while some of them can be seen as composite bulges with a classical component; however, none can be classified as a classical bulge. Auriga bulges show mostly an in-situ origin, $21\%$ of them with a negligible accreted fraction ($f_{\rm acc}$ < 0.01). In general, their in-situ component was centrally formed, with $\sim75\%$ of the bulges forming most of their stars inside the bulge region at $z=0$. Part of their in-situ mass growth is rapid and is associated with the effects of mergers, while another part is more secular in origin.
In $90\%$ of the Auriga bulges, the accreted bulge component originates from less than four satellites.  
We investigate the relation between the accreted stellar haloes and the bulges of the Auriga simulations. The total bulge mass shows no correlation with the accreted stellar halo mass, as in observations. However, the accreted mass of bulges tends to correlate with their respective accreted stellar halo mass. \\
\end{abstract}
\begin{keywords}
galaxies: formation -- galaxies: bulges -- methods: numerical 
\end{keywords}

\section{Introduction}
\label{sec:intro}

Milky Way (MW)-mass disc galaxies exhibit a large range of bulge properties and sizes, from prominent (e.g. M31), to almost non-existing bulges (e.g. M101). The diversity in the properties of these galaxies reveals the existence of different formation paths in the context of the current hierarchical galaxy formation paradigm \citep{WhiteRees1978}, which are not fully understood.
The study of bulges of MW sized galaxies in cosmological simulations is then an important task to try to explain the observed properties of the MW and luminous disc galaxies. 
    
    Galactic bulges are broadly classified as classical
bulges or pseudo-bulges. Historically, the classical bulges were defined as velocity dispersion dominated components in the center of disc galaxies. These objects have old stellar populations, exhibit a slow degree of rotation and present a spherical or elliptical shape.  Yet, many bulges show rotation, younger stellar populations and different features related to a disc-origin, like spiral structure, or nuclear bars. These differences have led to suggestions that such bulges (including bars) should be grouped and termed pseudo-bulges \citep[see][for a historical review on the subject]{KormendyKennicutt2004}.  
In the last decade, it has become increasingly clear that
a large fraction of disc galaxies in the local Universe hosts pseudo-bulges \citep{Gadotti2009, Kormendy2010, FisherDrory2011, FisherDrory2016, KormendyBender2018}. \citet{FisherDrory2011} found, in a volume limited sample within 11 {\rm Mpc},  that $80\%$ of disc galaxies with stellar masses larger than $10^{9} {\rm M_{\odot}}$ have pseudo-bulges or are bulgeless. Moreover, it has been determined in the last few years that MW-analogues commonly lack  a classical bulge component, or if there is one, it is not dominant. \citet{Kormendy2010} found that 11 out of 19 massive disc galaxies in the local Universe do not possess a classical bulge. In a recent paper,  \citet{KormendyBender2018} showed that two nearby MW-like galaxies (NGC 4565 and NGC 5746) host also a pseudo-bulge with similar properties to the MW-bulge and without signs of the presence of a classical bulge.
Observations and models of the MW itself indicate that a classical bulge component must be inconspicuous, if present \citep{Shen2010, DiMatteo2015, DeBattista2017, Gomez2018}.   Since, historically, classical bulges are considered to be formed in minor to intermediate mergers in a process analogous to that of the formation of elliptical galaxies \citep{Kauffmann1993}, this apparent low frequency of classical bulges in large disc galaxies have led some authors to claim a tension with the hierarchical clustering paradigm, where the formation of large galaxies should involve a relatively high amount of mergers.  However, analysis performed on single cosmological simulations, or small samples of them, suggest that the formation of pseudo-bulges is not infrequent. \citet{Guedes2013} used a simulation of a late-type galaxy, Eris \citep{Guedes2011}, to study the formation of its pseudo-bulge and found that most of the mass in the pseudo-bulge was formed in a bar configuration at high redshift that was later reshaped
into a dense flattened structure and inner bar. 
\citet{Okamoto2013} studied the formation channels of two pseudo-bulges in hydrodynamical resimulations of the Aquarius DM haloes \citep{Springel2008} and found that both of them formed at high redshift by means of the accretion of misaligned gas, with secular evolution contributing to less than $30\%$ of the final mass of the pseudo-bulges. \citet{DeBattista2018} studied in detail a high spatial and force resolution simulation from the FIRE project \citep{Wetzel2016} with signatures of {\it kinematic fractionation} \citep{DeBattista2017} during its bar evolution, a phenomenon that
can explain the observed properties of the MW bulge without the need of significant merger contributions to its formation. Under this scenario,  stellar populations with different kinematic properties at birth in barred galaxies end-up with different spatial distributions \citep[see also][]{Fragkoudi2017}. \citet{Buck2018a,Buck2018b} studied the inner region of a MW-like galaxy simulation and found two populations with distinct kinematics; one of them rapidly rotating and the other without significant rotation. Interestingly both populations formed mostly in-situ, with different initial angular momenta.  Despite these results already suggest that the formation of pseudo-bulges in simulated disc galaxies, within a $\Lambda$-CDM framework, is common, the frequency with which classical and pseudo-bulges form has not yet been addressed. This is due to the lack of a large and homogenous sample of simulated galaxies with enough resolution to study the detailed structure of the stellar component in the inner few kiloparsecs.  One of the goals of this paper is to survey the properties of bulges in one of the  largest samples of high-resolution hydrodynamical simulations of MW-mass galaxies, evolved in a cosmological context; namely the Auriga project \citep[][G2017 from now on]{Grand2017}. We wish to find the relative frequency of classical and pseudo-bulge formation.   

MW-analogs in the local Universe, where most of the bulges are found to be pseudo-bulges, present a
great diversity in its accretion histories, as revealed by detailed observations and determinations of properties of their stellar haloes \citep{Monachesi2016, Harmsen2017}, which is also seen in simulations \citep{DeSouza2018, Monachesi2018}.  It is not clear, however, under which conditions the accretion events that are involved in the build-up of the stellar halo contribute to the formation of the bulge, if they are involved at all \citep{Bell2017}.  The second goal of this paper is to study the origin and formation history of galactic bulges, and relate their properties to the properties of the corresponding stellar haloes  presented in \citet[][M2019 hereafter]{Monachesi2018}.
    This paper is organized as follows. In Section 2
we describe the simulations used in this work and present the definition of the Auriga stellar bulges.  In Section 3 we present the structural properties of the bulges and compare them with observational data. In Section 4 we study the formation history of the Auriga bulges, we define their in-situ and accreted
components and present the results and analysis for each component separately. We dedicate a subsection to study the relation of the Auriga bulges and their stellar haloes. 
Section 5 is devoted to the discussion of the results and Section 6 presents our summary and conclusions. 

\section{Methodology}
\label{sec:methodology}

\subsection{The Auriga simulations}
\label{sec:simulations}

The Auriga simulations consist of a suite of thirty high-resolution cosmological zoom simulations of the formation of galaxies in isolated MW-mass dark matter haloes, denoted throughout this paper by `AuN' with N varying from 1 to 30. 
The Auriga project was introduced in G2017 and we refer the reader to that paper for a detailed description of these simulations. Here, we briefly describe their main features.

Candidate haloes were selected from a lower resolution dark matter only cosmological simulation from the EAGLE project \citep{Schaye2015},
carried out in a periodic cube of side 100$h^{-1}$ {\rm Mpc}. A $\Lambda$-CDM cosmology was adopted, with parameters $\Omega_{m}  = 0.307$, 
$\Omega_{b}  = 0.048$,  $\Omega_{\Lambda}= 0.693$,  and  Hubble  constant  $H_{0}$  = 100 $h$ km s$^{-1}$
{\rm Mpc}$^{-1}$, $h=0.6777$ \citep{planck2013}. Gas was added to the initial conditions and 
its evolution was followed by solving the equations of ideal magnetohydrodynamics on an unstructured Voronoi mesh. Haloes were selected so that they satisfy: a) a narrow mass range of $1 < M_{200}/10^{12}\rm{M}_{\odot}<2$, comparable to that of 
the MW and b) an isolation criterion at  $z=0$,  placing each Auriga halo more distant than nine times its virial radius from any other halo of mass greater than 3\% of its own mass.  Each halo was re-simulated at higher resolution with the state-of-the-art N-body and moving mesh magnetohydrodynamics code {\sc arepo} \citep{Springel2010, Pakmor2016}. 
The typical mass of a dark matter
particle is $\sim 3 \times 10^{5}$ {\rm M}$_{\odot}$, and the baryonic mass resolution is 
$\sim 5 \times 10^4$ {\rm M}$_{\odot}$. The gravitational softening length of the stars and dark matter grows with the scale factor up to a
maximum of 369 {\rm pc}, after which it is kept constant in physical units. This value is large enough to resolve inner galactic regions.  As shown by GR2017, decreasing the softening lenght by a factor of 10 does not affect the overall properties of the resulting galactic models. The softening length of gas cells is scaled by the mean 
radius of the cell, with a maximum physical softening of 1.85 {\rm kpc} and is never allowed to drop below the stellar softening length. It is worth noting that, in high
density regions, gas cells are allowed to become smaller than the
gravitational softening length. This is particularly relevant for this study, where we will be 
focusing in the very inner regions of each simulated galaxy. However, it is important to keep in mind that our results could be sensitive to the subgrid physic model implemented in Auriga.

The simulations include a comprehensive model for 
galaxy formation physics which includes relevant baryonic processes, such as primordial and metal-line cooling \citep{Vogelsberger2013}; a sub-grid model for 
the interstellar medium that utilizes an equation of state representing a two-phase medium in pressure equilibrium \citep{Springel2003}; a model
for the star formation and stellar feedback that includes a phenomenological wind model \citep{Marinacci2014, Grand2017} and metal enrichment from SNII, SNIa and AGB stars \citep{Vogelsberger2013};
black hole formation and active galactic nucleus feedback \citep{Springel2005, Marinacci2014, Grand2017}; a spatially uniform, time-varying UV background after reionization at redshift six \citep{Faucher-giguere2009, Vogelsberger2013} and magnetic fields \citep{Pakmor2013, Pakmor2014}.
The model was specifically developed for the {\sc arepo} code and was calibrated to reproduce several observational results such as the stellar
mass to halo mass relation, galaxy luminosity functions and the history of the cosmic star formation rate density. 

The Auriga simulations reproduce a wide range of present-day observables, e.g. two-component disc-dominated galaxies with appropriate stellar masses, sizes, rotation curves, star formation rates and metallicities (G2017). The relatively large sample of thirty high-resolution simulations of late-type galaxies is an ideal set of simulations to study bulge formation in MW-sized galaxies and to relate our findings to both the accretion history of these galaxies and their stellar bulge properties. 
It is worth noting that the above described isolation  criterion  may bias the number and timing of encounters with 
satellite galaxies with respect to more dense environments. Late time mergers may be less common on average in the sample 
used in this paper, with respect to a random sample of Milky Way-sized DM haloes. However, these galaxies have not been 
specifically chosen to match the MW formation and merger history
(as in e.g. \citealt{BullockJohnston2005}) or the Local Group environment 
(as in the APOSTLE simulations of \citealt{Sawala2016}). Thus it is unclear how large this effect could be. 
We defer this analysis to a follow-up project based on the Illustris TNG50 simulations (Gargiulo et al. in prep.).

\subsection{Bulge definition}
\label{sec:bulge-def}

\begin{figure*}
\includegraphics[scale=0.55]{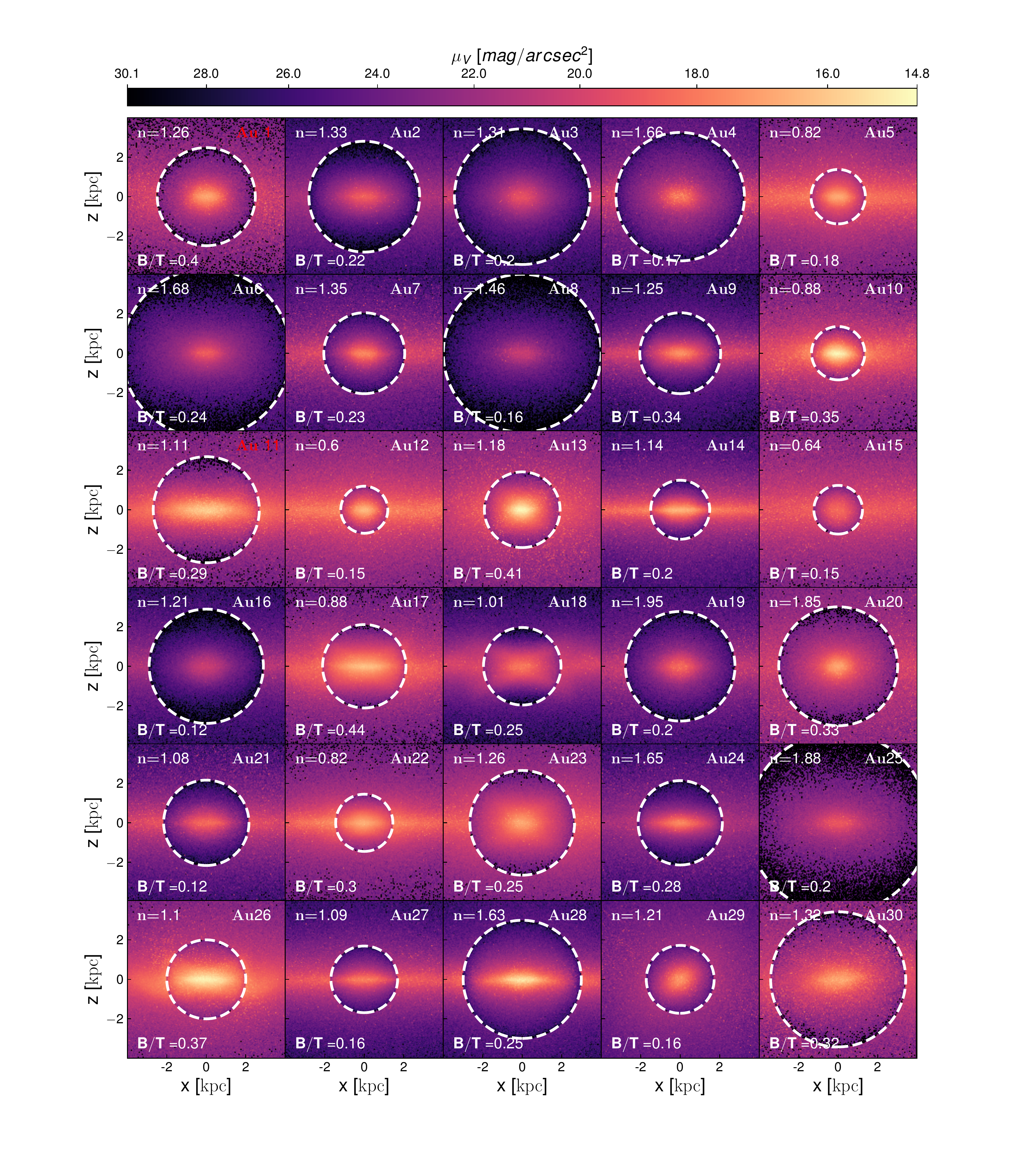}

\caption{Projected V-band surface brightness maps of the Auriga bulges shown edge-on at $z=0$ in a square region of 8 {\rm kpcs} on each side, centered in the galaxy. Bulges are defined using both an spatial and kinematic criteria (see text for details on the bulge definition in this work). The dashed white circle represents the bulge region within 2$\rm R_{\rm eff}$. Note that surface brightness beyond such circle is shown for all stellar particles, i.e. no kinematical cut. In the upper right corner of each panel the model name is shown. In the upper left corner the S\'ersic index and in the lower left corner the bulge-to-total ratio.}

\label{fig:bulges-Auriga}
\end{figure*}

Bulges of galaxies are defined in diverse ways in the literature, both in numerical simulations and observations. Observational studies often consider a spatial
definition, selecting a region surrounding  
the center of the galaxy to some extent \citep[see for example][for a MW bulge aimed survey]{Minniti2010}.
On the other hand, in numerical studies bulges are typically selected kinematically, to minimize the contribution of stars with orbits that are too circular, thus associated with the disc \citep[e.g.][]{Tissera2012,Guedes2013}. It is also common to limit the spatial extent of the bulge to separate it from the stellar halo component, even though this is not a physical criterion to determine where the bulge ceases to exist and gives place to the so-called inner halo. For example, \citet{Cooper2015} and M2019 choose a spherical region of 5 {\rm kpc} from the galactic center to mark the frontier between stellar halo and bulge. Other authors choose to avoid this spatial segregation, analyzing the overall ``central spheroids'' \citep{Tissera2018}. 

Here we use a combination of a spatial and a kinematical criteria.  We define the bulges of the Auriga simulations as all the stellar particles that fulfill the following two conditions at $z=0$. 
First, we consider the particles located inside a sphere of radius $r_{\rm bulge} = 2 R_{\rm eff}$, where $R_{\rm eff}$  is the
effective radius of the bulge and was derived from a S\'ersic profile fitting. The S\'ersic profiles are fitted together with an exponential profile (modeling the disc) to the face-on Auriga surface brightness profiles in the V-band derived using a non-linear least-square method. The full process is described in Appendix~\ref{sec:appendix}.

Second, we exclude stellar particles with pure disc kinematics. For that purpose, we 
consider only particles with circularities $\lvert \epsilon  \rvert  \leq 0.7$. The circularity parameter for  the stellar particles  at $z=0$ was calculated by G2017. It is defined as $\epsilon = J_z / J(E)$ \citep{Abadi2003}, where
$J_z$ is the angular momentum component perpendicular to the disc plane of a star particle and $J(E)$ is the maximum possible angular momentum for the orbital energy, {\it E}, for the same particle.  The median of the mass
removed from the region defined as the bulge due to this circularity cut, for all Auriga simulations, is $m_{f_{{\rm \epsilon > 0.7}}} = 0.19 $. Only for Au25 the total mass removed within $2 R_{\rm eff}$ reaches almost $45\%$. The total mass removed from each galaxy in this bulge defined region due to the circularity cut is listed in Table~\ref{table:bulgeprop}, along with other derived properties of the Auriga bulges. It is important to note that, when comparing with observations, the bulge is defined as closely as possible as it
is done in the particular observational analysis. The goal is to make fair comparisons with the observed quantities in each particular case.

\section{Structural Properties of Bulges}
\label{sec:generalprop}

In Fig.~\ref{fig:bulges-Auriga} we show the
edge-on projected surface brightness maps in the V-band of the stellar particles that form the bulges. 
These were obtained by computing the photometry of each stellar particle, which represents a single stellar population, using \citet{BruzualCharlot2003} population synthesis models.
In the corners of each panel, we show the simulation name,
the S\'ersic index, and the bulge-to-total stellar mass ratio (${\rm B/T_{sim}}$).\\
The bulges show an interesting diversity in morphology. While some of them show a more rounded or elliptical morphology, there are some clear peanut/boxy (p/b) or X-shaped bulges like Au17, Au18 or Au26. Perhaps the most interesting example is Au18, which shows an X-shape that resembles the morphology of the MW bulge \citep{McWilliamZoccali2010, Nataf2010}. We note here that taking a more restrictive limit in the circularity cut in the definition of galactic bulges (e.g. $\lvert \epsilon  \rvert  \leq 0.6$ instead of $\lvert \epsilon \rvert  \leq 0.7$) does not significantly affect our results. Disc-like features of bulges are present even with a more stringent circularity threshold.
Understanding the origin of this diversity is one of the aims of this study and a series of subsequent papers.

The S\'ersic index was extracted from the same profile
used to define the effective radius,  discussed in
the Appendix~\ref{sec:appendix}. ${\rm B/T}$ was computed in two different ways. First, we computed ${\rm B/T_{sim}}$ as the ratio between the total mass of the stellar particles inside the bulge region as defined in the previous section and the total mass of the stellar particles inside a sphere of 0.1 $\times$ the virial radius of the host. In addition, we estimated the ${\rm B/T_{v}}$ from the two-component fit described in Appendix~\ref{sec:appendix}. We integrated the fitted S\'ersic function and divided the result by the integral of the sum of the S\'ersic function and the exponential function describing the disc. The resulting ${\rm B/T}$ for all simulations using each method are shown in Table~\ref{table:bulgeprop}.
We find that for all of the Auriga bulges ${\rm B/T_{sim}} < 0.5$ and  most of the Auriga bulges, ${\rm B/T_{v}} < 0.5$, which is a common threshold above which the presence of a classical bulge is ensured, in observational studies \citep{Kormendy2015, BrooksChristensen2015}. However, in some cases, observed bulges can be classified as classical even if the galaxy shows a low {\rm B/T} \citep{FisherDrory2011}.  Au13 and Au26  show values of luminosity-weighted ${\rm B/T_{v}} > 0.5$. One of the reasons for this result is that these simulated galaxies experienced high levels of star formation in the last snapshots of the simulation, (See the star formation histories in Fig.~\ref{fig:insitu-tvsr-1} and~\ref{fig:insitu-tvsr-2} in Sec.~\ref{sec:insitu}). In the case of Au13, the surface brightness profile shows a prominent bump due to the bar that was extracted during the fitting procedure, but high levels of bar contamination in the light profile remain. Light profile decompositions of barred Auriga galaxies adding a third component for the bar show {\rm B/T} below 0.5 for Au13 and Au26 (Blazquez-Calero et al., private communication).
The effective radii of bulges vary from
$R_{\rm eff}=0.6 {\rm kpc}$ for Au12, to 
$R_{\rm eff}=2.29 {\rm kpc}$ for Au25.  The S\'ersic indexes have values between $n=0.6$ for Au12 and $n = 1.88$ for Au15. 
A list of bulge parameters presented here can be found in Table~\ref{table:bulgeprop}.

\begin{figure}
\includegraphics[width=0.46\textwidth]{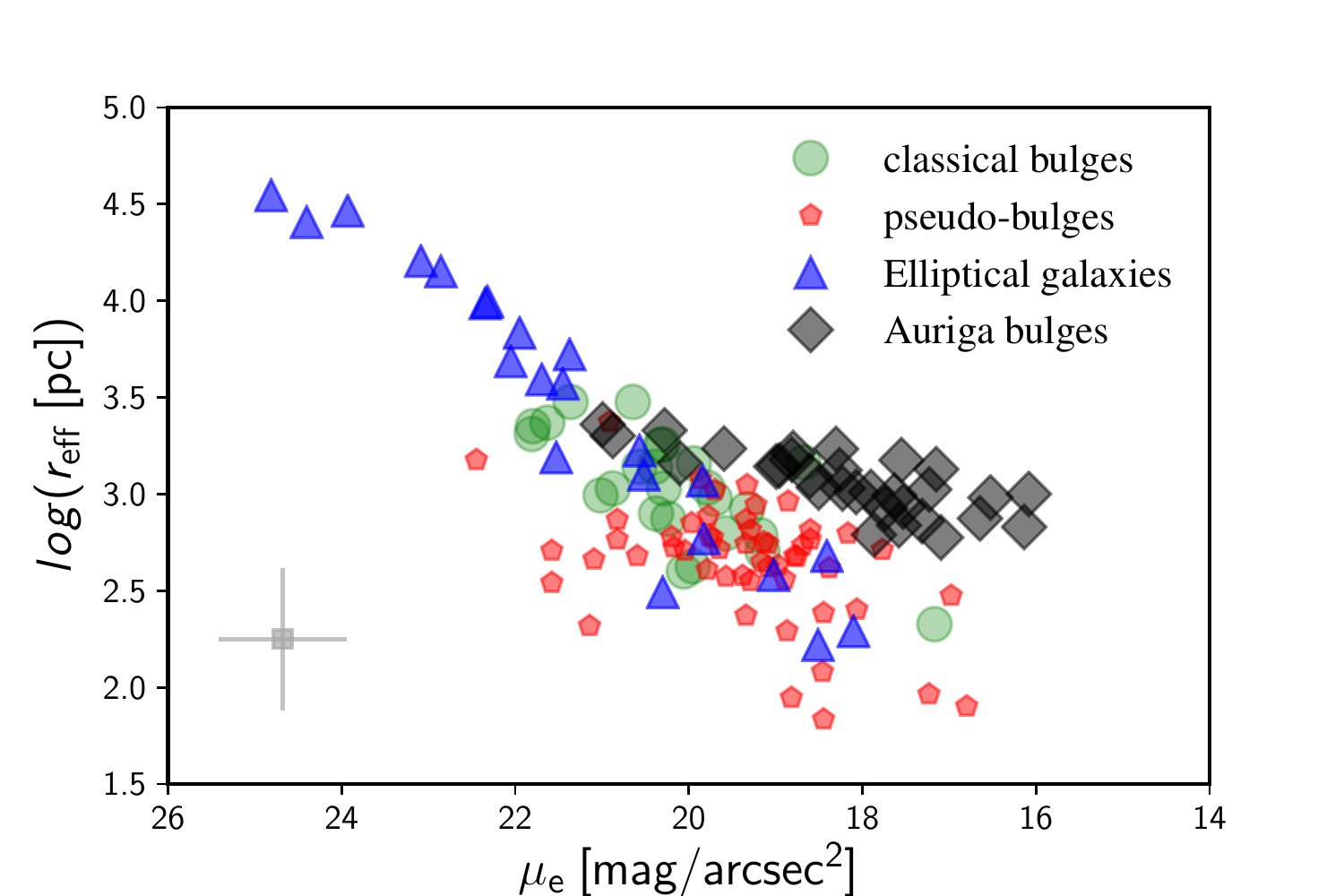}
\includegraphics[width=0.46\textwidth]{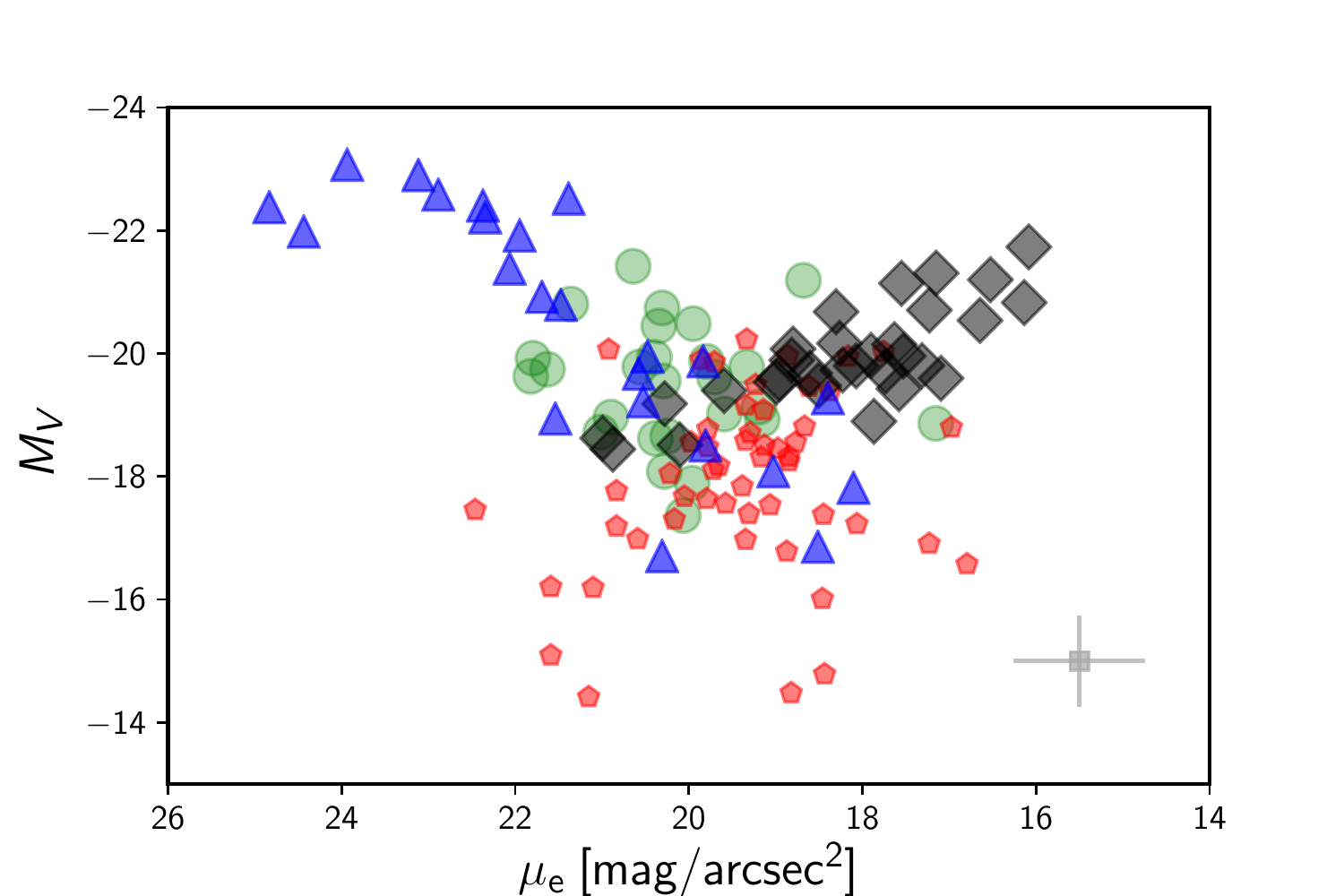}
\includegraphics[width=0.46\textwidth]{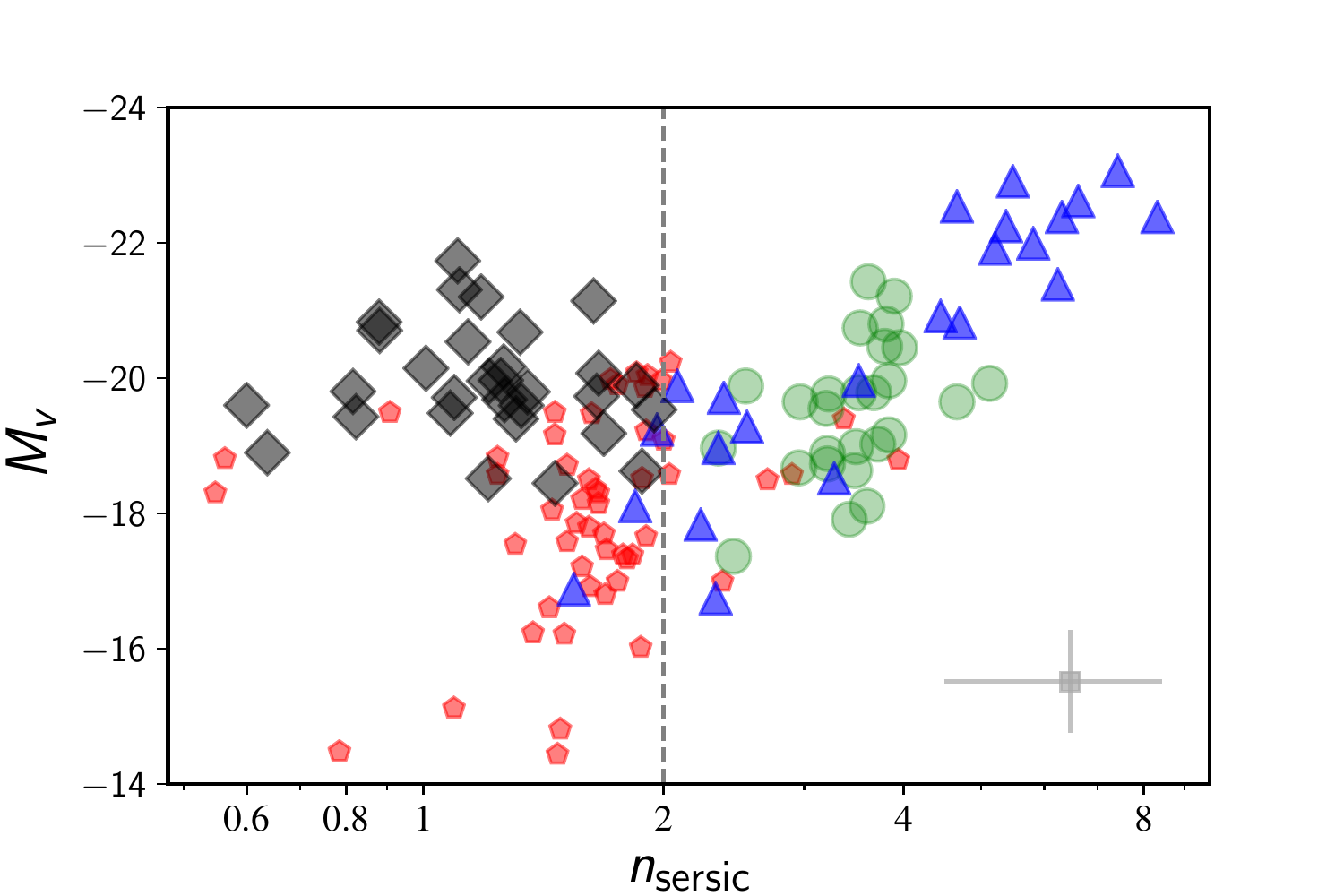}

\caption{{\it Top:} Effective radii of galactic bulges as a function of surface brightness at effective radii in the V-band. The Auriga bulges are indicated with black diamonds. Red pentagons, blue triangles, and green circles are observational data from \citep{FisherDrory2008}. The Gray square with error bars represents the averaged errors of the observations. {\it Middle:} Absolute magnitude as a function of surface brightness at the effective radius in the V-band. {\it Bottom:} Absolute magnitude as a function of sersic index. The limit $n=2$ is indicated with a dashed vertical line. All the Auriga bulges show $n < 2$. }
\label{fig:scalingrel}
\end{figure}

The photometric classification of observed galactic
bulges as classical or pseudo-bulges can be difficult. The most straightforward approach is using the S\'ersic index, which has been shown to correlate with bulge type. 
\citet{FisherDrory2008} studied the bulges of 79 spiral galaxies observed with the {\it Hubble Space Telescope} (HST) spanning a range of morphologies from early-type to late-type. They found a bi-modal distribution of S\'ersic index and showed that more than $90\%$ of the bulges that are classified as pseudo-bulges morphologically by visual inspection, have S\'ersic indices $n \lesssim 2$. On the other side, classical bulges have commonly S\'ersic indices $n>2$.  The dependence of this structural parameter with bulge type was already suggested in previous studies, such as \citet[][and references therein]{KormendyKennicutt2004}. But it was also proven that using the S\'ersic index as the only parameter of bulge classification can be too simplistic.  \citet{FisherDrory2011} used a combination of S\'ersic index, morphological classification by visual inspection, and star formation activity to discriminate between bulge types and found that for composite bulges (those which show a distinguishable spheroidal classical component and, at the same time, show pseudo-bulge morphological features such as central spiral patterns, a ring or bar), the use of structural 
parameters as S\'ersic index are not reliable. 

If we only consider the S\'ersic index parameter to classify the bulges, we find that all of the Auriga bulges should be classified as pseudo-bulges, i.e. $n < 2$.  Yet, as previously highlighted \citep[see also][]{Kormendy2015} a multiparameter classification must be carried out to reduce the error in bulge classification. In the following we analyze the scaling relations, intrinsic shapes and degree of ordered rotation of the Auriga bulges.

\begin{figure*}
\includegraphics[scale=0.41]{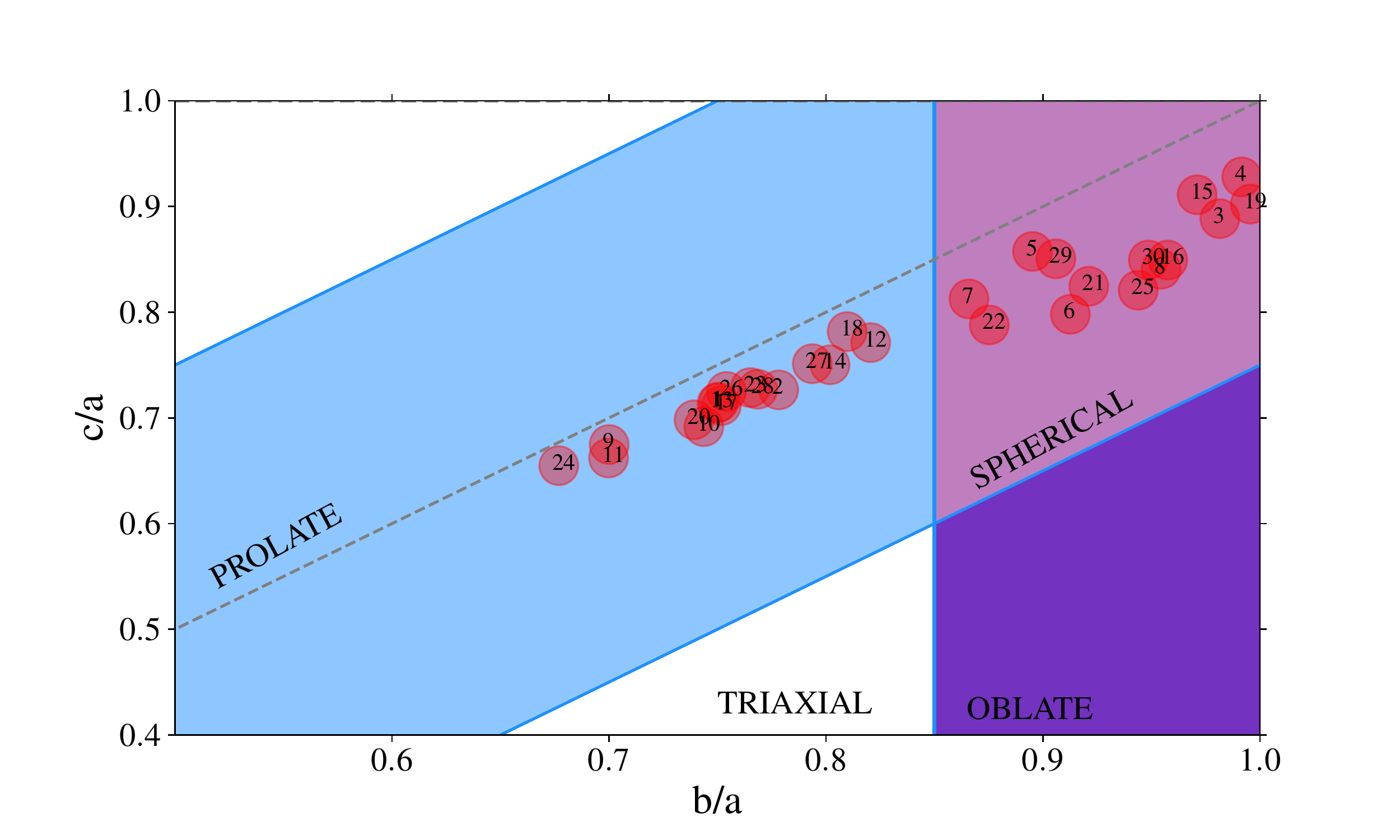}
\includegraphics[scale=0.42]{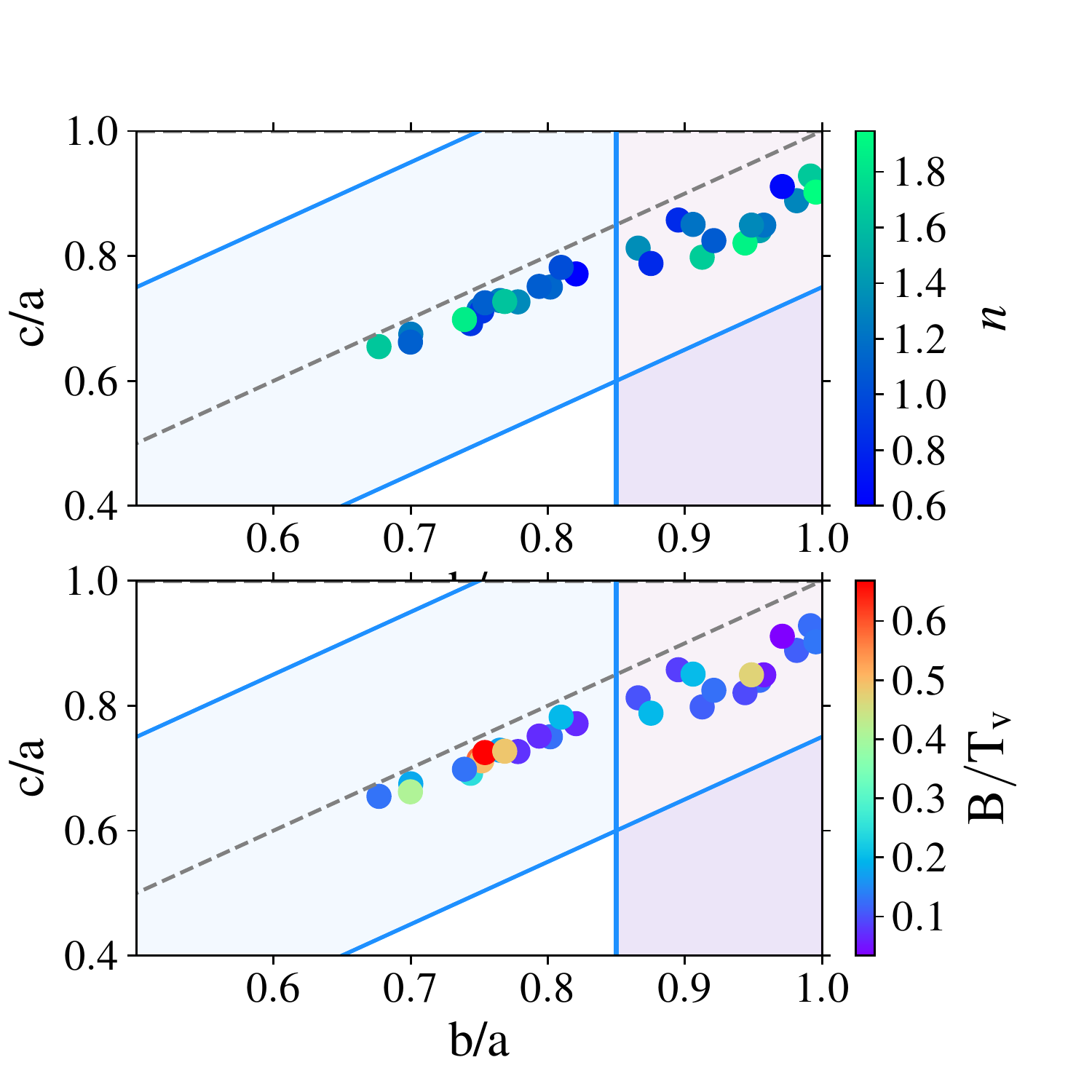}

\caption{{\it Left panel:} Intrinsic axial ratios c/a vs. b/a for
the Auriga bulges. The shaded areas in violet, pink and light blue represent the oblate, spherical and prolate regimes, respectively, defined in
\citet{Costantin2018a}. {\it Right panels:} The same diagram as in
the left panel, but with bulges coloured by S\'ersic index (top) and bulge-to-total ratio (bottom).}

\label{fig:coaboadiagram}
\end{figure*}

\subsection{Scaling relations}

Scaling relations of bulges can be used as a complementary tool to help in the bulge classification.  \citet{Gadotti2009} used the relation between effective radius and surface brightness at the effective radius of elliptical galaxies, known as the Kormendy diagram \citep{Kormendy1977}, as another way to determine the bulge type. They assumed that all galaxies outside $1 \sigma $ of the relation followed by elliptical galaxies can be considered pseudo-bulges, but again \citet{FisherDrory2011} showed that many pseudo-bulges selected morphologically can lie in the same region than classical bulges on this diagram and that this criterion can not be used in isolation to determine a single bulge type. Yet, pseudobulges and classical bulges do show different behaviours in this diagram when seen as different samples.  
In Fig.~\ref{fig:scalingrel} we show scaling relations of the bulges for the Auriga galaxies.
In the upper panel, we show the Kormendy diagram. In the middle panel, we present the relation between the absolute magnitude in the V-band and the surface brightness at the effective radius and in the bottom panel, we present the relation between absolute magenitude and sersic index. The Auriga bulges are indicated with black diamonds. Observational data from \citet{FisherDrory2008} of classical bulges, pseudo-bulges and elliptical galaxies are shown as green circles, red pentagons, and blue triangles, respectively. Average error bars of the observed data are indicated in a lower corner of the panels. 
To make a fair comparison with observations, in these diagrams all the quantities are derived directly from the 
2-component decomposition of the V-band surface brightness profiles (See Appendix~\ref{sec:appendix}), without applying the kinematic cut previously defined in
Sec~\ref{sec:bulge-def}.  Hence, $M_V$ and $\mu_e$ values also include the contribution from particles with circular orbits 
inside the effective radius.  It is expected that classical bulges follow the relation found for ellipticals, 
but pseudo-bulges usually show a larger scatter in these diagrams. In the Kormendy diagram (upper panel), 
the Auriga bulges occupy a rather narrow range in effective radii and  appear to be systematically larger than 
the observed pseudo-bulges. While some simulated pseudo-bulges have similar sizes to the observed ones, the high surface
brightness objects ($\mu_{e} < 17~ \rm{mag ~ arsec}^{2}$) tend to be larger than observed.  In general, we find that the Auriga bulges show a correlation in this diagram, with larger effective radii for smaller surface brightness. However, the slope shown by the Auriga bulges in the $\log{(r_{\rm eff})}~$--$~\mu_{\rm{eff}}$ relation is flatter than the slope shown by the observed clasical bulges or elliptical galaxies. Observed pseudo-bulges in this diagram show little to no-correlation.

In the middle panel of Fig.~\ref{fig:scalingrel} we show the $M_{\rm V}~$--$~\mu_{\rm{eff}}$ relation. While elliptical galaxies and classical bulges follow the same trend in this space, i.e. more luminous ellipticals and classical bulges are less centrally concentrated, the bulges of Auriga follow an opposite trend. The observed pseudo-bulges in this panel only show a large scatter. However, the results for the extended sample of galaxies with pseudo-bulges presented in \citet{FisherDrory2011} follows qualitatively the same trend seen for the Auriga bulges, i.e. the more luminous the denser, as highlighted by the authors of that work (see their Fig.~8)\footnote{Results in this paper are not quantitatively compared with observations in \citet{FisherDrory2011} because they used images with filters centered $3.6\mu$ from the Spitzer telescope, which are not comparable with the available magnitudes in the Auriga simulations.}. A group of Auriga bulges ($23\%$ of the total), namely Au10, Au11, Au13, Au14, Au17, Au26, Au28 , show to be more luminous than any observed pseudo bulge and present a higher surface brightness than most observed pseudo and classical bulges of this sample. In the bottom panel of Fig.~\ref{fig:scalingrel}, we show the relation between  $M_{\rm V}~$ and sersic index. We can see that the sersic index of observed classical bulges follow broadly the relation found for elliptical galaxies and observed pseudo bulges show a large scatter with no signs of correlation. Although a group of the Auriga bulges show absolute magnitudes that are comparable with those of classical bulges, as already shown in the $M_{\rm V}~$--$~\mu_{\rm{eff}}$ relation, their sersic index and total magnitude are not correlated. \\
There are two points to take into account in the comparison with these set of observations.
i) \citet{FisherDrory2008} exclude from the
analysis the central regions of the surface brightness profiles when
they identify a nuclei and due to resolution limits we cannot identify
substructure in the underlying peak of surface brightness produced
by the bulge component (See Appendix~\ref{sec:appendix}). ii)  Additionally, extinction by dust is neglected in the simulations.  Because of these two caveats, the total magnitudes of the Auriga bulges may be overestimated compared with the observed ones. Besides, this group of galaxies with higher abosulute magnitudes in the V-band are all actively forming stars in the last snapshots of the simulation (as can be seen in Fig.~\ref{fig:insitu-tvsr-1} and Fig.~\ref{fig:insitu-tvsr-2}. As disscused in Sec.~\ref{sec:discussion}, simulated stellar feedback might not suppress enough star formation in this set of simulations, thus likely generating more massive simulated bulges.
It is worth also noting that the observational sample contains the whole range of disc galaxy morphologies and, most likely, a wider range in DM halo masses. Our simulations are restricted to a narrow range of DM haloes ($1-2 \times 10^{12} {\rm M}_{\odot}$). Note also that, as shown by \citet{FisherDrory2011}, a reliable classification of a bulge into either pseudo or classical cannot be done only with the object position in the Kormendy diagram. Instead, all available diagnostics, such as morphology,  S\'ersic index, and kinematics should be combined when possible.

\subsection{Intrinsic shapes}
\label{sec:intrinsic-shapes}

Different bulge formation mechanisms are thought to contribute to shape bulges in different ways. 
\citet{Costantin2018b} stated that the intrinsic shape of bulges provides a complementary classification between classical and pseudo-bulges. Their results are based on a statistical method to derive bulge intrinsic shapes from the observed projected 2D shape of galaxies \citep{Mendez-Abreu2008, Mendez-Abreu2010, Costantin2018a}. Here we compare the results of our simulations with the conclusions drawn from these observations.

We quantify the intrinsic 3D shapes of bulges by computing the eigenvalues of their mass distribution inertia tensor, adopting the same approach as M2019. We compute the principal axes of the mass distributions in the whole bulge, considering the center of the mass distribution as the most bound DM particle.
Fig.~\ref{fig:coaboadiagram} shows the c/a vs b/a diagram used by \citet{Costantin2018a} to analyze correlations between the intrinsic shapes and properties of bulges. They defined different regimes in this diagram, which are colour coded and indicated with text in Fig.~\ref{fig:coaboadiagram}. Auriga bulges occupy two
well defined loci. 53$\%$ of the Auriga bulges show a very clear prolate shape, while the remaining 47$\%$ are clustered in the region defined as the spherical regime. We find that the two major axes of the inertia tensor are well aligned, or have very low inclinations, with respect to the disc plane. Galaxies with bulges in the prolate regime are barred and the major axis of the mass tensor of bulges is in the direction of the bar major axis. 

In the right panels, we show the same diagram, but with symbols colour coded according to the S\'ersic index (top) and ${\rm B/T_{sim}}$ ratio (bottom). We can see that bulges with higher S\'ersic indexes tend to cluster in the spherical regime, with values of  $0.8 \lesssim {\rm c/a} \lesssim 0.9$ and $0.9 \lesssim {\rm b/a} \lesssim 1$. Two bulges exhibit values of ${\rm b/a} \approx 1$. The bulges with moderately high S\'ersic indexes present more spherical shapes, as found by \citet{Costantin2018a}. Bulges with lower S\'ersic indexes tend to show more prolate shapes, although a group of them  also occupies the spherical regime. This is in disagreement with the results of \citet{Costantin2018a}, since they find that the low S\'ersic index bulges are more commonly triaxial. Interestingly, none of our bulges show triaxial shapes. 
In the right bottom panel we see that  bulges with larger values of ${\rm B/T_{v}}$ are also prone to occupy the prolate regime. This is at odds with the results shown by  \citet{Costantin2018a}, who found that bulges with larger ${\rm B/T_{v}}$ tend to be oblate systems. One possible reason for this discrepancy is that they probe the full range of disc morphologies, ranging from lenticulars to late type spiral galaxies. Here we are narrowing the analysis to disc galaxies in MW-sized haloes, which are expected to have similarities in their formation mechanisms, unlike the broad sample selected in the CALIFA survey.  This discrepancy can also arise because \citet{Costantin2018a} define {\it a priori} the orientation of the axes in the plane of the observed projected bulge, while we define the major, intermediate and minor axes according to the mass distribution.
A detailed analysis of the projected and intrinsic shapes of the bulges in high resolution simulations with a broader range of DM halos masses would be of great interest to constrain the connection between bulge shapes and their formation processes, and to shed light on this apparent discrepancy.

\subsection{Degree of ordered rotation}
\label{sec:rotation}

The degree of ordered rotation of bulges is a fundamental quantity
to discriminate between bulges and pseudo-bulges. The seminal work by \citet{Kormendy1982} showed that bulges present kinematical properties that differentiate them from elliptical galaxies. They used the ${\rm V_{max}}/\sigma$ diagram \citep{Illingworth1977}, where ${\rm V_{max}}$ is the maximum velocity in the line-of-sight (LOS), typically estimated  using long-slit spectroscopy along directions parallel to the galaxy major axis, and $\sigma$ is the velocity dispersion in the LOS. Classical bulges usually have some degree of ordered rotation and oblate shapes, while elliptical galaxies are supported by velocity dispersion and show anisotropy. 
When classical bulges coexist in barred disc galaxies, for example, they can acquire angular momentum from the bar \citep{Saha2015}. For its part, pseudo-bulges usually show a higher degree of ordered rotation \citep{Kormendy1993, KormendyKennicutt2004} as their formation is linked to the discs and bars.

With the advent of integral-field spectroscopy (IFS) surveys like SAURON \citep{deZeeuw2002}, ATLAS$^{\rm 3D}$ \citep{Cappellari2011} or MaNGA \citep{Bundy2015}, detailed kinematical structure in galaxies can be analyzed. In this context, \citet{Binney2005} proposed another way for computing $V/\sigma$ that could take 
into account all the available observed data. \citet{Cappellari2007} used this
for the first time, applied to the SAURON survey. We follow also this implementation and compute $V/\sigma$ as:  

\begin{equation}
\left(\frac{V}{\sigma}\right)_e^2\equiv\frac{\langle V^2 \rangle}{\langle\sigma^2 \rangle} = \frac{\sum_{n=1}^{N} F_n\, V_n^2}{\sum_{n=1}^{N} F_n\, \sigma_n^2}
\label{eq:vsigma}
\end{equation}

\noindent where the index n denotes the {\it nth}-pixel of 0.2 squared {\rm kpc} size in the edge-on view of Auriga galaxies inside 1 effective radius, a typical radial extent probed in the observations. For this calculation all  stellar particles within one effective radius, including the disc, are taken into account, as done in observations. $V_n$ and and $\sigma_n$ are the mean velocity in the direction perpendicular to the $x-z$ plane and  the velocity dispersion, respectively,  on the n-pixel. $F_n$ is the corresponding total flux. Ellipticity is obtained from the mass tensor analysis explained in Sec.~\ref{sec:intrinsic-shapes} as:

\begin{equation}
(1- \varepsilon)^2= \frac{\langle y^2 \rangle}{\langle x^2 \rangle} = \frac{\sum_{n=1}^{N} F_n\, y_n^2}{\sum_{n=1}^{N} F_n\, x_n^2},
\label{eq:eps}
\end{equation}

\noindent where $x$, and $y$ are the main axes of the mass tensor of the region inside 1 effective radius.

\begin{figure}
\includegraphics[scale=0.57]{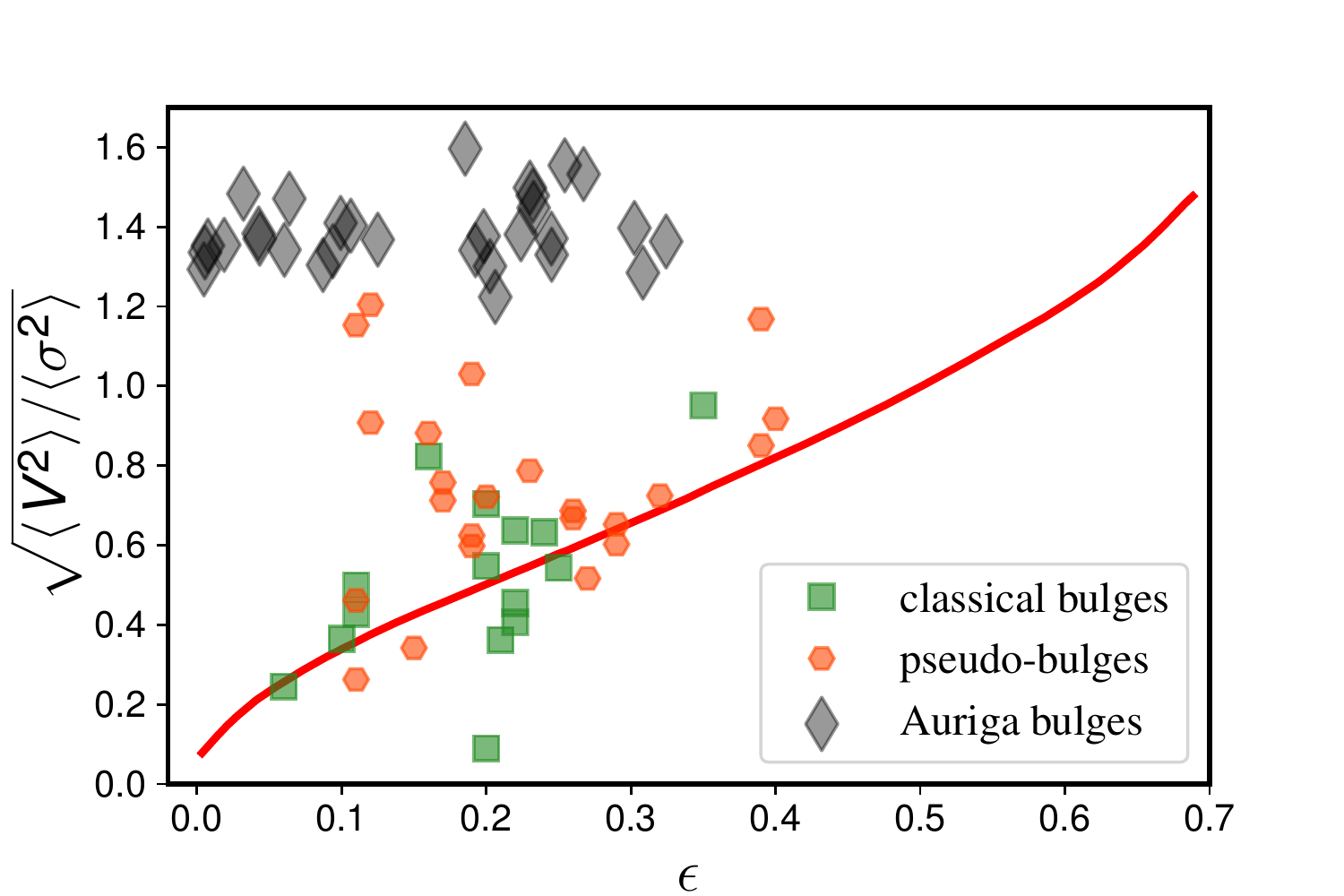}

\caption{{\it Left panel:} ($V/\sigma$, $\epsilon$) diagram for our sample of Auriga bulges shown in black diamonds. Red hexagons and green squares are observational data from \citet{Fabricius2012}. The red line indicates the oblate line that describes oblate spheroids that are isotropic and flattened only by rotation.}

\label{fig:vosigma-elip}
\end{figure}

In Fig.~\ref{fig:vosigma-elip} we show the $V/\sigma$ diagram for the Auriga bulges.
They all show strong signatures of rotation in a moderate range of ellipticities.
The oblate line is indicated with a red line. This line is approximated as $\epsilon/(1-\epsilon)^{1/2}$ and describes oblate spheroids that are isotropic and flattened only by rotation \citep{Binney1978, Kormendy1982}. Classical bulges in this diagram usually occupy positions near the oblate line or below and pseudo-bulges show usually a higher degree of ordered rotation. To illustrate this we also plot in Fig. ~\ref{fig:vosigma-elip} observational data from \citet{Fabricius2012}. Auriga bulges show a higher degree of ordered rotation than the observed pseudo-bulges with a higher degree of ordered rotation found in  \citet{Fabricius2012}, reaching values of $V/\sigma > 1.5$. It is worth noting that we compute the LOS velocity with galaxies oriented perfectly edge-on. This is not the case for these observations, that were selected with inclinations low enough to classify bulges morphologically and velocities were later corrected for the inclination effect.  The rotation degree shown by central regions of the Auriga simulations is a strong indication of the prevalence of pseudo-bulge formation in this sample of simulated galaxies. 

\begin{table*}
\caption{Table of bulge parameters at $z=0$. The columns are 1) Model name; 2) Bulge stellar mass, defined as the sum of all stellar particles inside the bulge region defined in Sec.~\ref{sec:bulge-def}; 3) S\'ersic Index; 4) Bulge-to-total ratio computed as the ratio between the sum of masses of particles inside bulges as defined in Sec.~\ref{sec:bulge-def} and the sum of the masses of all stellar particles inside $0.1\times R_{\rm vir}$; 5) Bulge-to-total ratio derived from two-component fits to the surface brightness profile in the V-band  6) Bulge effective radius; 7) Accreted bulge fraction; 8) Minor-to-major axis ratio c/a; 9) Intermediate-to-major axis ratio b/a; 10) Percentage of in-situ bulge stars formed outside the bulge region; 11) Fraction of in-situ (all) stars younger than 8 Gyrs; 12) Fraction of stars inside the bulge region with circularities $\epsilon$ > 0.7}

\centering
\begin{tabular}{c c c c c c c c c c c c c}
\hline
\hline
Sim &       
$\frac{M_{\rm bulge}} {[\rm 10^{10} {\rm M}_{\odot}]}$ &
$ n_{\rm v}$  &
$ {\rm B/T_{sim}}$  &
$ {\rm B/T_{\rm v}}$ &
$ \frac{r_{\rm eff_v}} {[\rm kpc]} $  &
$ f_{\rm acc}$ &
$ {\rm c/a} $ &
$ {\rm b/a} $ &
$f_{\rm >2Reff}[\%] $ &
$f_{\rm <8 Gyr} $ &
$f_{\rm \epsilon > 0.7} $ \\
 \hline
 \hline
 
 Au1 & 1.01 & 1.26 & 0.37 & 0.24 & 1.24 & -- & 0.73 & 0.77 & -- & -- & 0.25 &\\
 Au2 & 1.36 & 1.32 & 0.19 & 0.07 & 1.40 & 0.15 & 0.72 & 0.78 & 33.1 & 0.37 (0.36) & 0.15  &\\
 Au3 & 1.55 & 1.30 & 0.20 & 0.11 & 1.72 & 0.23 & 0.89 & 0.98 & 17.9 & 0.11 (0.09) & 0.32 &\\
 Au4 & 1.53 & 1.65 & 0.21 & 0.12 & 1.62 & 0.42 & 0.92 & 0.99 & 42.9 & 0.93 (0.81) & 0.25 &\\
 Au5 & 1.31 & 0.82 & 0.19 & 0.09 & 0.68 & $< 0.01$ & 0.86 & 0.90 & 42.8 & 0.52 (0.51) & 0.23 &\\
 Au6 & 1.00 & 1.68 & 0.21 & 0.11 & 2.13 & 0.14 & 0.81 & 0.94 & 31.5 & 0.38 (0.33) & 0.36 &\\
 Au7 & 1.01 & 1.35 & 0.21 & 0.10 & 1.02 & 0.20 & 0.81 & 0.87 & 36.3 & 0.94 (0.79) & 0.20 &\\
 Au8 & 0.58 & 1.46 & 0.19 & 0.12 & 2.00 & 0.15 & 0.84 & 0.94 & 40.9 & 0.31 (0.33) & 0.24 &\\
 Au9 & 2.01 & 1.25 & 0.33 & 0.18 &1.02 & 0.03 & 0.67 & 0.70 & 41.2 & 0.46 (0.45) & 0.15 &\\
 Au10 & 1.97 & 0.88 & 0.33 & 0.25 &0.67 & $< 0.01$ & 0.68 & 0.73 & 56.8 & 0.74 (0.74) & 0.15 &\\
 Au11 & 1.83 & 1.10 & 0.26 & 0.41 & 1.34 & -- & 0.65 & 0.69 & -- & -- & 0.16 &\\
 Au12 & 0.83 & 0.60 & 0.14 & 0.07 & 0.60 & 0.01 & 0.76 & 0.81 & 53.1 & 0.71 (0.71) & 0.16 &\\
 Au13 & 2.10 & 1.18 & 0.34 & 0.53 & 0.95 & 0.01 & 0.71 & 0.74 & 51.1 & 0.79 (0.78) & 0.18 &\\
 Au14 & 1.88 & 1.14 & 0.18 & 0.12 & 0.75 & 0.02 & 0.75 & 0.80 & 46.1 & 0.74 (0.72) & 0.18 &\\
 Au15 & 0.27 & 0.64 & 0.07 & 0.03 & 0.62 & 0.02 & 0.88 & 0.99 & 52.1 & 0.84 (0.82) & 0.24 &\\
 Au16 & 0.65 & 1.20 & 0.12 & 0.05 & 1.44 & 0.01 & 0.85 & 0.96 & 31.9 & 0.18 (0.18) & 0.35 &\\
 Au17 & 3.41 & 0.88 & 0.45 & 0.49 & 1.06 & $ < 0.01$ & 0.71 & 0.75 & 51.6 & 0.58 (0.58) & 0.17 &\\
 Au18 & 1.95 & 1.01 & 0.24 & 0.19 & 0.98 & $ < 0.01 $ & 0.78 & 0.81 & 57.4 & 0.52 (0.52) & 0.16 &\\
 Au19 & 1.07 & 1.94 & 0.20 & 0.13 & 1.39 & 0.22 & 0.90 & 0.99 & 32.3 & 0.71 (0.69) & 0.24 &\\
 Au20 & 1.35 & 1.85 & 0.28 & 0.13 & 1.51 & 0.18 & 0.72 & 0.77 & 33.0 & 0.32 (0.38) & 0.14 &\\
 Au21 & 1.03 & 1.08 & 0.13 & 0.13 & 1.08 & 0.10 & 0.82 & 0.91 & 38.2 & 0.74 (0.69) & 0.26 &\\
 Au22 & 1.90 & 0.81 & 0.31 & 0.20 & 0.72 & $ < 0.01$ & 0.80 & 0.89 & 51.5 & 0.60 (0.60) & 0.19 &\\
 Au23 & 2.35 & 1.26 & 0.26 & 0.17 & 1.33 & 0.03 & 0.73 & 0.76 & 47.6 & 0.52 (0.50) & 0.15 &\\
 Au24 & 1.90 & 1.65 & 0.29 & 0.13 & 1.06 & 0.01 & 0.65 & 0.67 & 29.8 & 0.28 (0.28) & 0.15 &\\
 Au25 & 0.55 & 1.88 & 0.17 & 0.09 & 2.29 & 0.08 & 0.83 & 0.97 & 32.2 & 0.30 (0.28) & 0.44 &\\
 Au26 & 4.10 & 1.10 & 0.37 & 0.62 & 1.00 & 0.11 & 0.72 & 0.75 & 51.8 & 0.77 (0.68) & 0.22 &\\
 Au27 & 1.52 & 1.09 & 0.16 & 0.07 & 0.84 & 0.05 & 0.75 & 0.79 & 37.9 & 0.32 (0.31) & 0.17 &\\
 Au28 & 4.03 & 1.63 & 0.38 & 0.42 & 1.49 & 0.24 & 0.75 & 0.80 & 51.8 & 0.87 (0.76) & 0.19 &\\
 Au29 & 1.67 & 1.21 & 0.18 & 0.20 & 0.86 & 0.28 & 0.84 & 0.91 & 34.9 & 0.52 (0.50) & 0.16 &\\
 Au30 & 1.26 & 1.32 & 0.29 & 0.47 & 1.71 & 0.16 & 0.84 & 0.96 & 40.7 & 0.80 (0.74) & 0.36 &\\
\hline
\hline
  \vspace{-0.3cm}
\\

\end{tabular}
\label{table:bulgeprop}
\end{table*}

\section{Formation history of the Auriga Bulges}

In this section, we study the main physical mechanisms behind the formation of the bulges in the Auriga simulations. A fundamental step to understand the diversity in their formation history  is 
to characterize the origin of the stellar particles that populate each of them.

\subsection{Accreted vs In-Situ components}
\label{sec:AccInsituDef}

One possible and common practice to define their origin is to subdivide star particles according to whether they formed {\it in-situ} or were accreted into the bulge from satellite galaxies \citep[e.g.][]{Zolotov2009, Tissera2012, Pillepich2015}. Studies in the literature have subtle differences in these definitions and some gray areas can arise. For example, stars formed in a starburst during a merger event from gas coming from the satellite have been considered as both an accreted or in-situ origin. Stars formed in tidal tails can cause the same ambiguity too. 

\begin{figure*}
\includegraphics[scale=0.46]{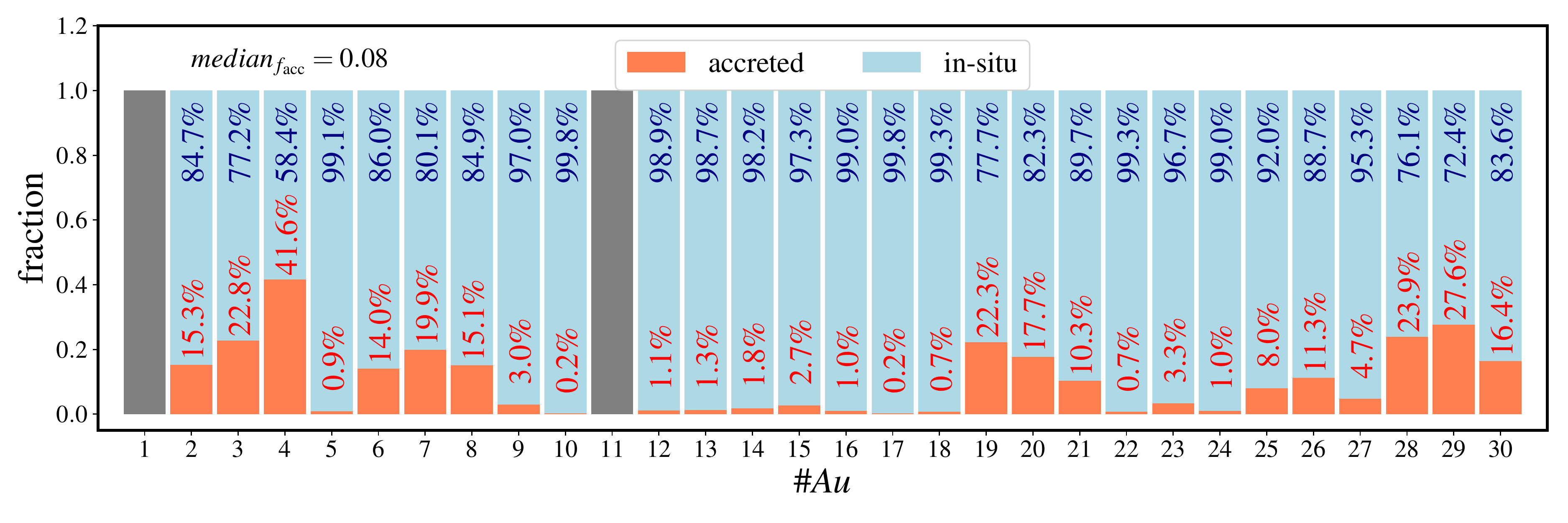}

\caption{Fraction of accreted and in-Situ stars in bulges of the {\rm Auriga} galaxies. Despite the rich merger history of the Auriga galaxies, all its bulges are dominated by in-situ formed stars. The median accreted fraction is $f_{\rm acc} = 0.08$. Moreover,  $23\%$ of the simulated galactic bulges show negligible contributions from an accreted component. As indicated in the text Au 1 and Au 11 are excluded from this analysis because of ongoing merger activity.}

\label{fig:Bar-acc-insitu}
\end{figure*}

Here, we define the accreted component of bulges 
as the stellar particles born bound to satellites of the host galaxy, that are members of the host bulge at $z=0$ (see the definition of Auriga bulges in Sec.~\ref{sec:bulge-def}).
On the other hand, the in-situ component of
the Auriga stellar bulges is defined as all $z=0$ bulge stellar particles 
that were born inside the virial radius of the host galaxy and were not bound to any distinct satellite at the time of formation. This definition includes the star particles formed from the stripped gas accreted during a satellite accretion.  
In Fig.~\ref{fig:Bar-acc-insitu} we show the fractions of bulge stars that belong to the accreted and in-situ components.  Inside the bars, we show the fractions in terms of percentages. Au1 and Au11 bars are grey, because we exclude them from the current analysis. The reason for this is that both Au1 and Au11 show an ongoing major accretion event, which makes it difficult to define the different components of these galaxies.  Only Au4 shows a considerable accreted bulge fraction of 0.42. In all the other cases, the accreted fractions are below 0.28 and for a few bulges like Au5, Au10, Au17, Au18 and Au22, the percentage of accreted material is less than $1\%$. Despite the rich merger histories of many of our simulated galaxies (see e.g. M2019), the bulge accreted fractions are generally marginal. 

\begin{figure*}

\includegraphics[width=0.35\textwidth]{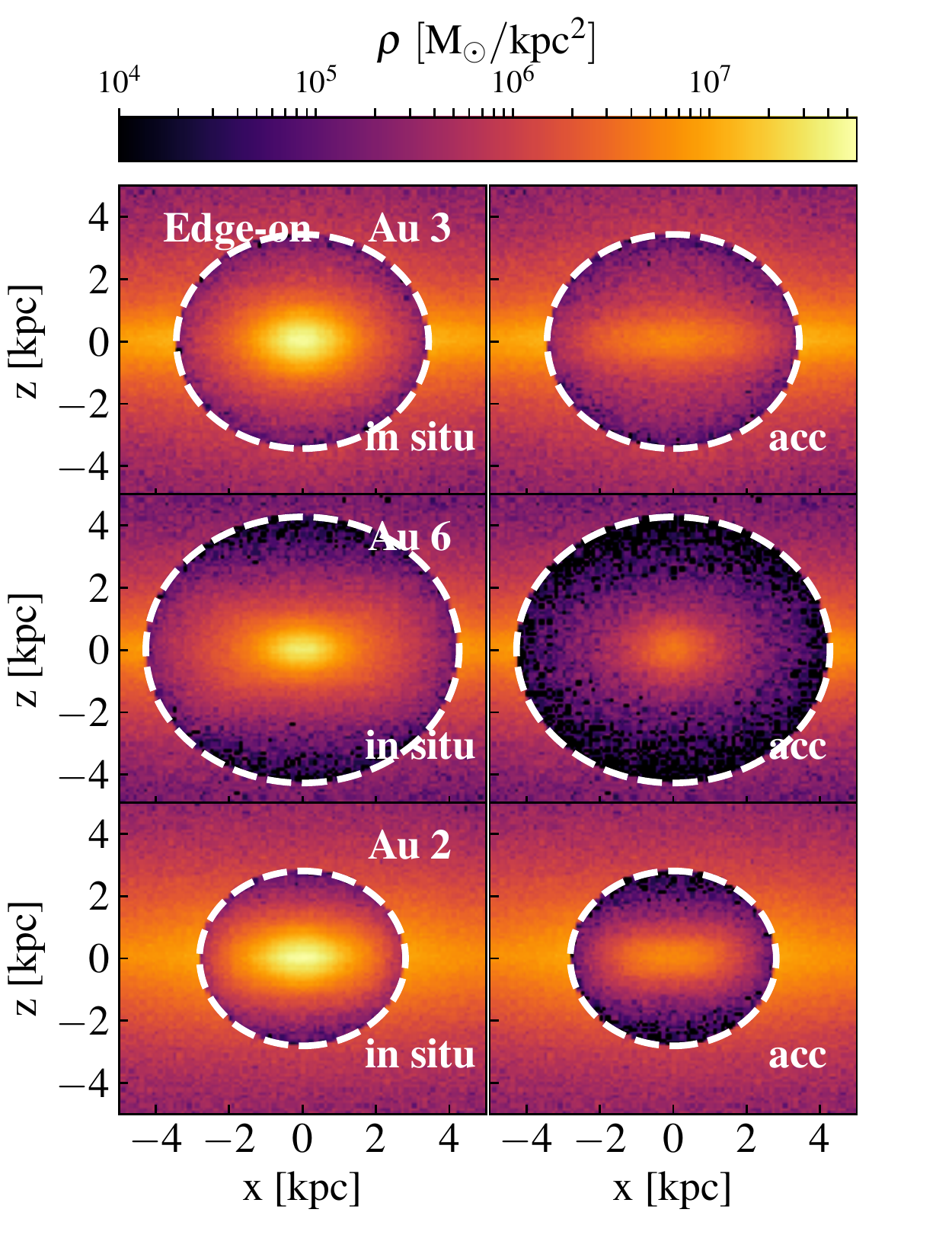}
\includegraphics[width=0.35\textwidth]{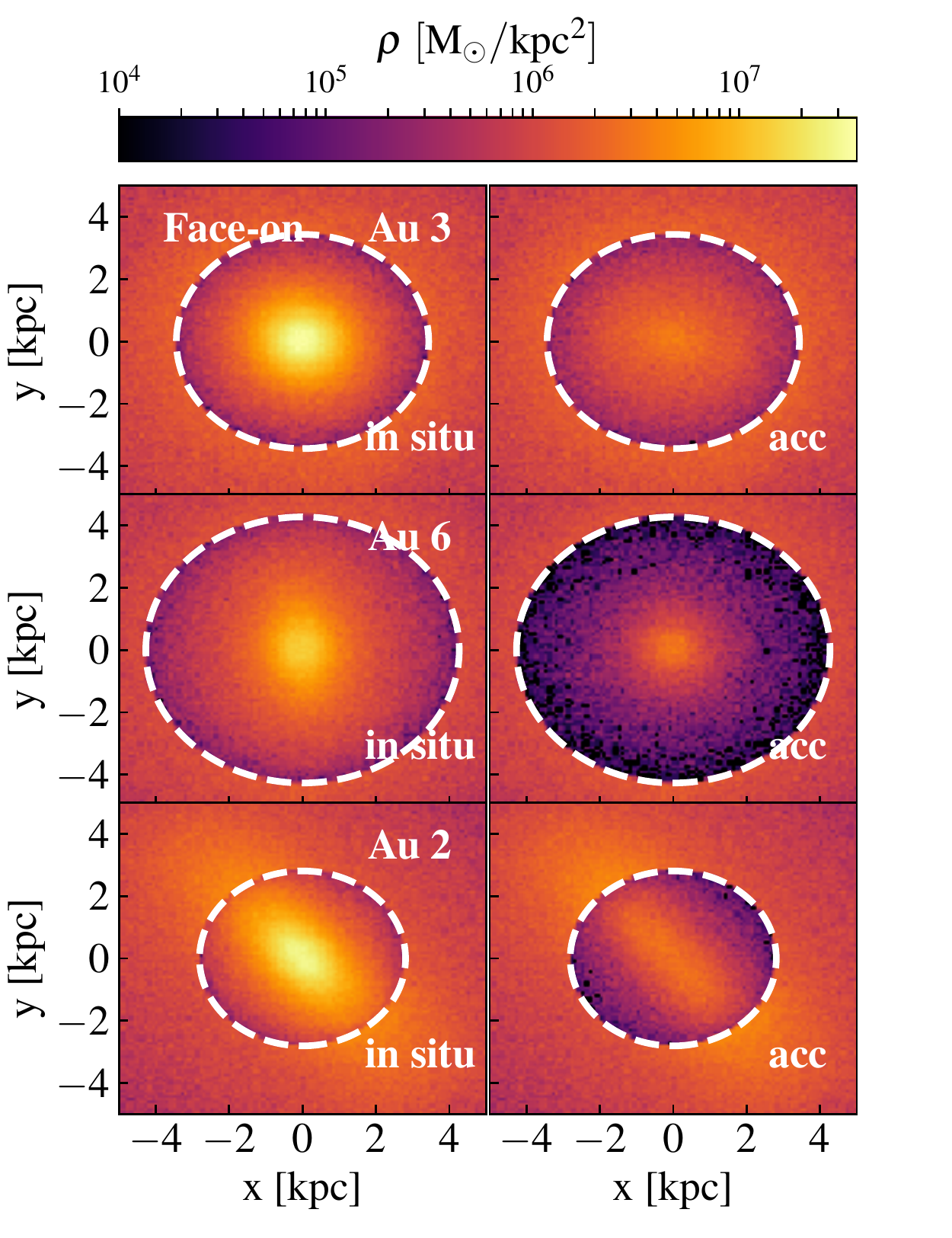}
\includegraphics[width=0.245\textwidth]{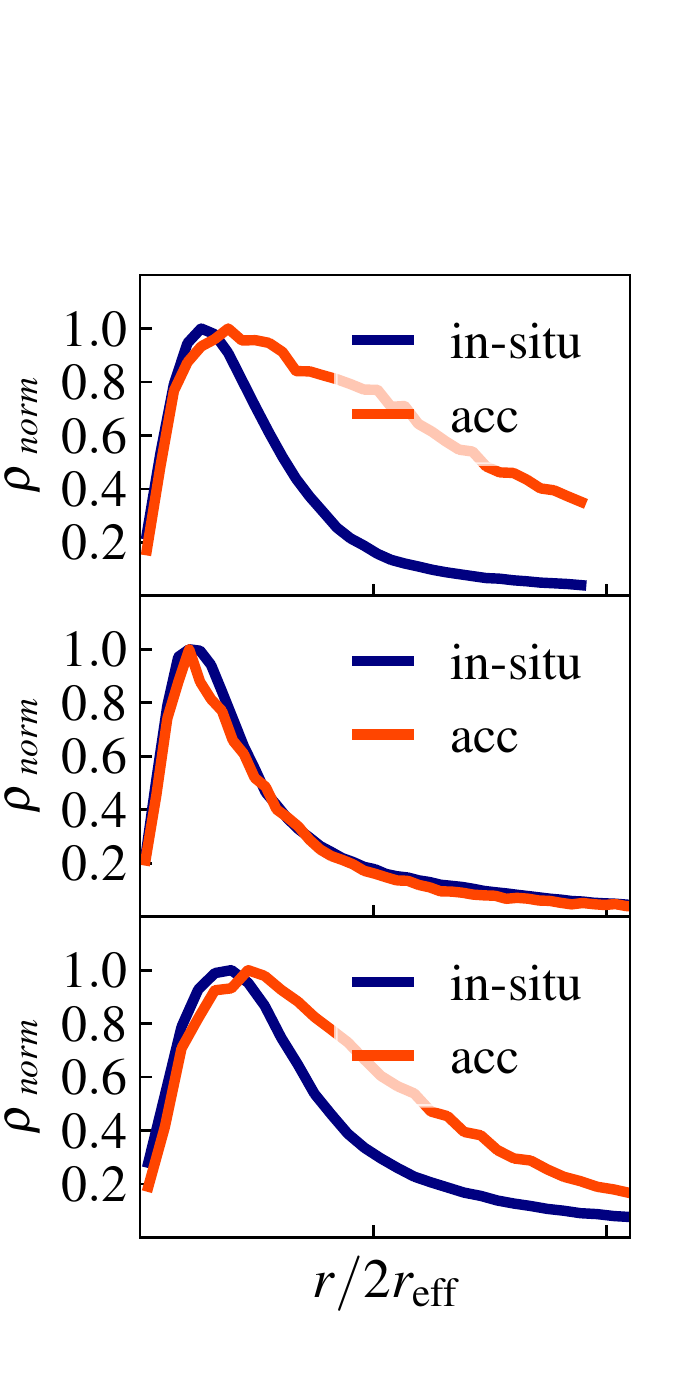}

\caption{{\it Left panels:} Projected density maps of the in-situ and accreted components of bulges of a subsample of the Auriga galaxies as seen edge-on.  The dashed white circle represents the bulge region within 2$\rm R_{\rm eff}$.  
{\it Middle panels}: Same as in left panels, but as seen face-on.
{\it Right panels:} Normalized spherical density profiles of the in-situ and accreted components as indicated in the legends. Au2 presents bar-shaped in-situ and accreted components. Au3 shows an accreted component less concentrated than the in-situ component. Au6 shows the same level of concentration in both components.}
\label{fig:AccIns-DensShape1}
\end{figure*}

In the left and middle panels of Fig.~\ref{fig:AccIns-DensShape1} we show the mass density maps of the in-situ and accreted bulge components of a subsample of Auriga bulges as seen edge-on and face-on, respectively. In the right panels, we show the mass density profiles of both components, normalized to their corresponding maximum value. We choose three examples that depict three different types of behaviour of the accreted and in-situ bulge components.
We find that the in-situ components are more centrally concentrated than the accreted counterparts in 14 bulges ($\sim 50\%$), as shown for Au3 in the first panel. The second panels show the example of Au6. Here the normalized spherical density distribution of the bulge accreted component follows almost exactly the one shown by in-situ bulge. Simulated bulges that show this behaviour ($\sim 11\%$ of Auriga bulges) have coincidentally a high fraction of accreted bulge stars. The similarity in the shape of the
distribution could be the consequence of the violent relaxation of the stellar particles formed in-situ, together with the accreted particles of a massive satellite. Indeed, the morphology of both components of Au6 shows a striking similarity in the left panels. 
Interestingly, for 4 out of 28 bulges ($\sim 14\%$) the accreted component follows the disc-like shape of their in-situ component when seen edge-on (see \citealt{Gomez2017} for a dedicated paper regarding the ex-situ discs in the Auriga simulations), and the bar signature when seen face-on. The case of Au2 shown in the third panel is the clearest example where a bar-like distribution is seen in the accreted component.
The remaining bulges ($\sim 25\%$ of Auriga bulges) have too low accreted fractions to make a meaningful comparison between the spatial distribution of both components, but we note that in all these cases of low accreted bulge mass fractions, the accreted bulge components are also less concentrated than the in-situ components.
This result indicates that it is possible for the accreted stellar particle distribution to be reshaped by internal secular processes after being accreted \citep[e.g.][]{Saha2012}. The existence of this phenomenon in real galaxies could help to explain to some extent the apparent underabundance of classical bulges in disc galaxies. 

\subsection{Formation history of the in-situ component}
\label{sec:insitu}

\begin{figure*}
\includegraphics[scale=0.42]{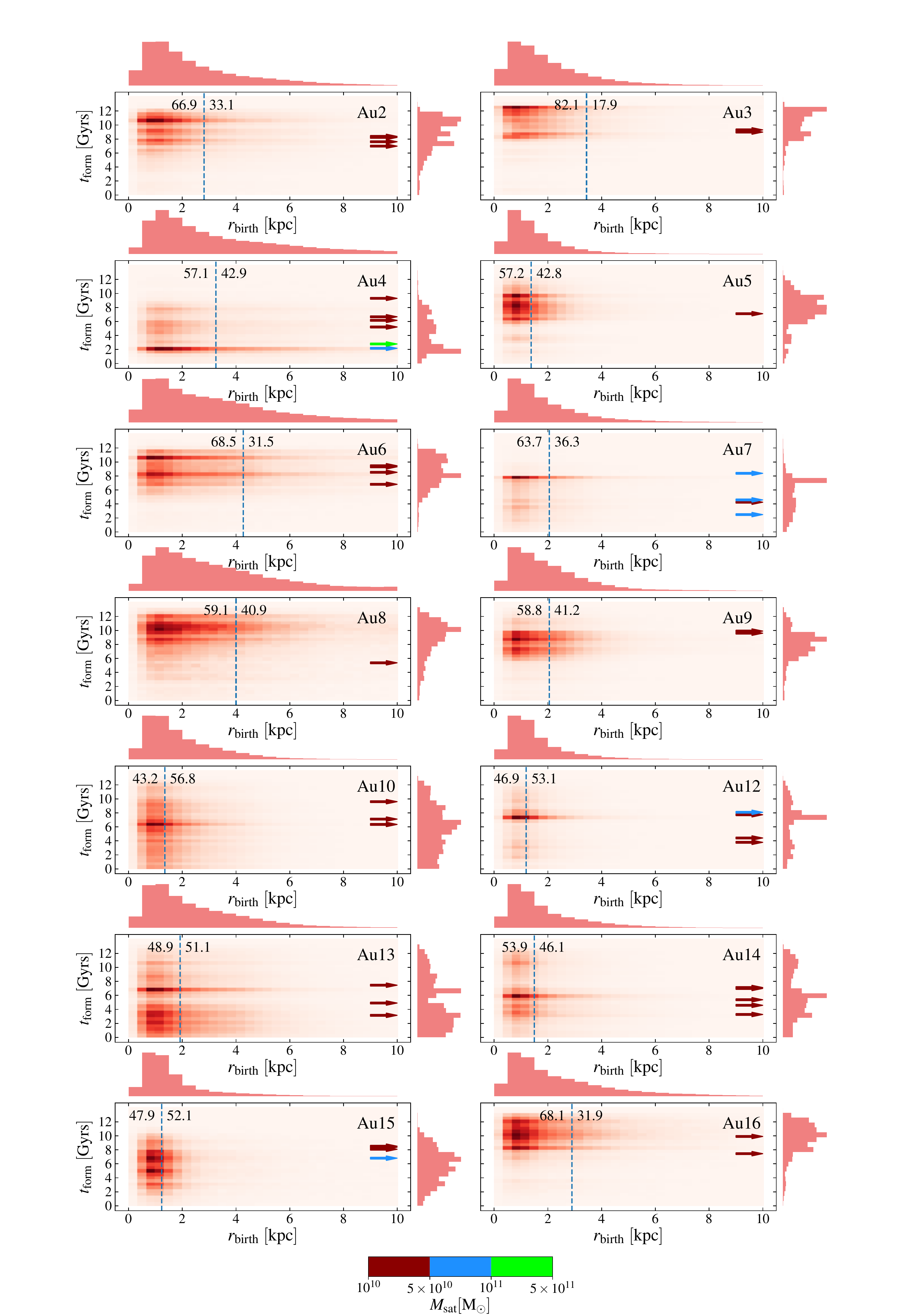}

\caption{Formation times as a function of birth radii of stellar particles inside the bulge region at $z=0$, for each of the first 14 Auriga simulations. Normalized distributions of birth radii and formation times are shown above the horizontal axis and to the right of the vertical axis, respectively. The vertical blue dashed lines indicate the spatial limit of the bulges at $z=0$. The values to the left and right of the dashed lines indicate the fraction of in-situ bulge stars formed inside and outside the bulge region, respectively. Mergers times are indicated by arrows that are coloured according to the total mass of the merged satellite, as indicated by the color bar. Only mergers with satellites of total masses above $10^{10} {\rm M}_{\odot}$  that had occurred in the last 10 {\rm Gyrs} of the simulation are shown in this figure.}

\label{fig:insitu-tvsr-1}
\end{figure*}

\begin{figure*}
\includegraphics[scale=0.42]{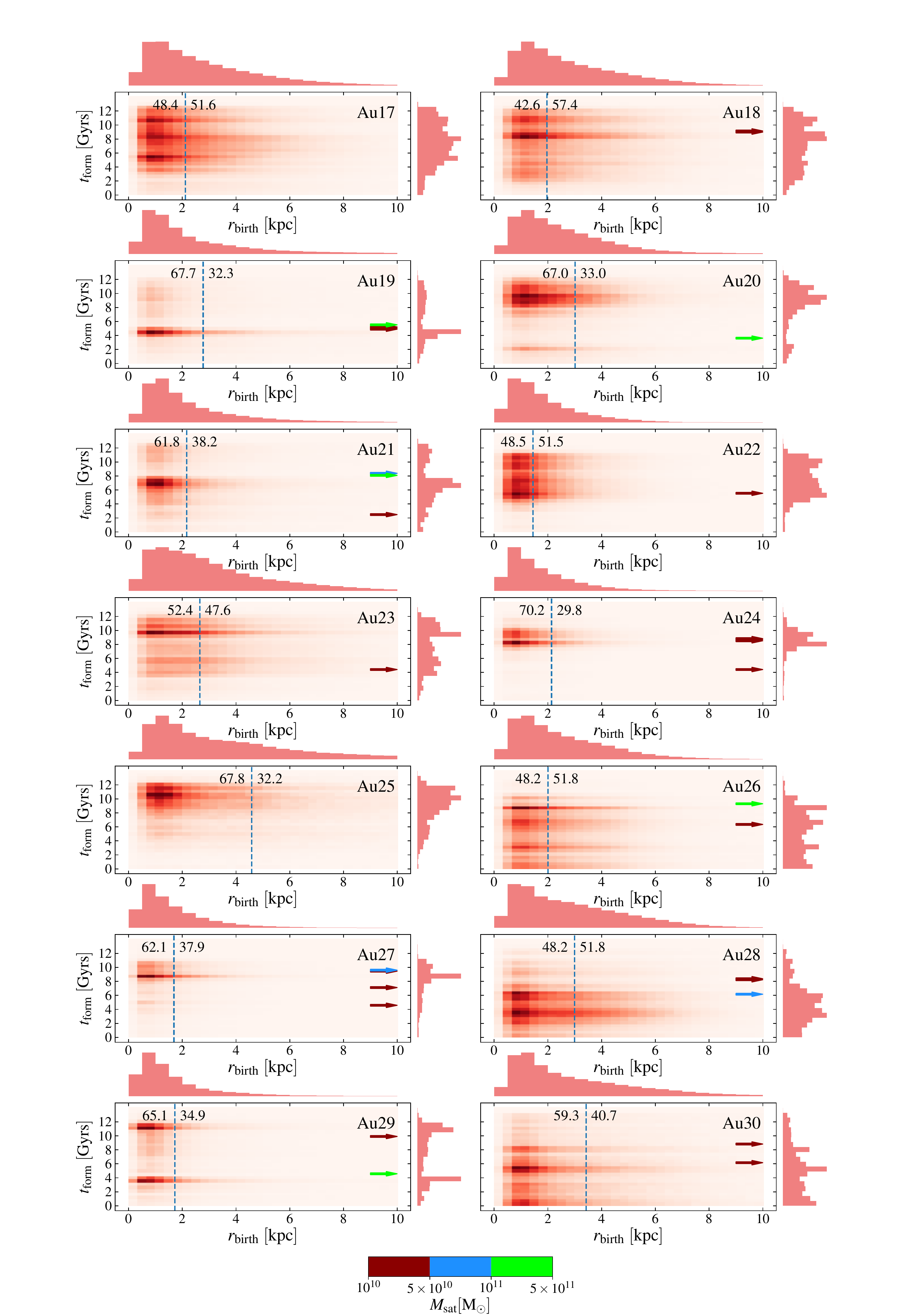}

\caption{As in Figure \ref{fig:insitu-tvsr-1} for the rest of the 14 Auriga simulations. }

\label{fig:insitu-tvsr-2}
\end{figure*}

The simplest question that we can ask about the in-situ population is where and when did it form. We show in Fig.~\ref{fig:insitu-tvsr-1} and Fig.~\ref{fig:insitu-tvsr-2} the stellar particles formation time (age) distribution, $t_{\rm form}$, of the $z=0$ in-situ bulge, as a function of birth radius, $r_{\rm birth}$. Here $r_{\rm birth}$ is defined as the galactocentric distance of each particle at the nearest snapshot of their formation.  The blue dashed line on each panel represents the spatial extent of the bulge region. The percentages of stars formed inside and outside this limit are shown next to left and right of this line, respectively. In all cases, the birth radii distributions show that the star formation is peaked towards the central regions of the galaxy. This is in line with observations using the ALMA telescope that show high levels of star formation in the central regions of disc galaxies at high redshift ($z \sim 2$) \citep{Tadaki2017}. We can see that, although star particles are formed predominantly inside the bulge regions defined at $z=0$, there are several galaxies which formed a large fraction of their in-situ bulge outside this region. It is interesting to note the cases of Au10, Au12, Au13, Au15, Au17, Au18, Au22, Au26 and Au 28 which have an extended distribution and a dominant fraction ($f^{>2{\rm reff}}_{r_{\rm birth}} > 0.50$) of bulge stars that were formed outside the bulge region. 

The bulge formation time distributions on the right axis of Fig.~\ref{fig:insitu-tvsr-1} and Fig.~\ref{fig:insitu-tvsr-2} show great diversity and, in many cases, a high fraction of intermediate-age and young stars ($t_{\rm form} < 8~ {\rm Gyr} $). The fraction of these younger stars in the in-situ bulge, indicated for each simulation in Table~\ref{table:bulgeprop},  range from 0.11 for Au3 up to 0.94 for Au7. We can see star formation histories with peaks at different redshifts.  During these
peaks of star formation, the birth radii span a large range, sometimes reaching 10 {\rm kpc}.
Typically, these peaks are associated to mergers with satellite galaxies. The most relevant merger events  are highlighted with arrows on each panel. The colors of the arrows indicate the total masses of the merging satellites, as indicated in the color bar. We only show mergers with satellites of total masses above $10^{10} {\rm M}_{\odot}$  that had occurred in the last 10 {\rm Gyrs} of the simulation. Interestingly, some of the peaks on the bulge star formation histories are associated with groups of small mergers instead of mergers with large
satellites. However, not all peaks are  correlated with galaxy mergers. For example Au17 shows several peaks of star formation
with no associated merger. In many of those cases, we found that different forms of interactions due to flybys of massive satellites are triggering the star formation in the host galaxy \citep[see e.g.][]{2016MNRAS.456.2779G}. During the mergers,
the host discs are strongly perturbed and star formation is enhanced. Part of these newly formed stars settle down into the central regions as the system relaxes to an equilibrium.
Additionally, a fraction of stars is formed in central starbursts driven by gas funneling in the case the merger is
wet \citep{Hopkins2009, Bustamante2018}.
Noticeably, the diversity shown in the star formation histories is shown to be related to the expected diverse merger histories
experienced by the galaxies during their evolution in a cosmological scenario.  

On the other hand, some of the simulations show extended star formation periods, which are associated with secular evolution processes within the galactic discs. The model Au18  provides a typical example of these processes. Au18 undergoes a merger with a satellite in the period between $8-9~{\rm Gyrs}$ ago, which is shown by the arrow and generates a strong $t_{\rm form}$ peak in Figure~\ref{fig:insitu-tvsr-2}. As shown by \citet{Grand2016}, who characterized the vertical heating of the Auriga discs as a function of time, Au18 develops a strong bar after this merger event (see their Figure 5). As a result, the star formation rate 
inside the bulge radius is increased by the funnelling of cold gas which loses angular momentum due to the torques exerted by the bar. Studies based on detailed hydrodynamic simulations have shown that the accretion of gas with low angular momentum in the inner galactic region produces off-axis shocks that drive gas into internal nuclear rings where stars are formed \citep{SandersHuntley1976, Kim2012, Kim2018}. Other galaxies like Au17 and Au22 also show extended periods of star formation inside and outside the bulge region that are a consequence of secular evolution. These bulges develop strong non-axisymmetric instabilities during its evolution. In addition, direct cold gas accretion  can also play a role in the enhancement of gas density and subsequent star formation both inside and outside the disc region \citep[e.g.][]{Sales2012}. Au18 shows a significant fraction of bulge stellar particles that formed outside the bulge region after the merger event. These are particles that can be brought to the bulge due to an interplay of processes. It is well known that stars in discs can migrate from the
original position where they were born in the galactic disc. Dynamical processes that are responsible for such behaviour are the presence of transient spiral
arms, which change the angular momentum of stellar particles near the corotation
radius and drive outward and inward streaming motions of these particles, and the resonant coupling between the bars and the spiral patterns.
\citep{SelwoodBinney2002, Minchev2010, Grand2012}. 
The same physical process can occur due to a bar, that can exert the loss of angular momentum to the particles captured in resonances. This way, particles can be caught inside the bulge radius during their orbit in the last snapshot of the simulation. 
Quantifying the evolution of bulges and the precise relative contributions of each mechanism is beyond the scope of this paper and is postponed to future work.
The relative importance of these mechanisms of in-situ
bulge growth give place to the scatter that is seen in the Auriga bulges properties, together with the different accretion histories.

\subsection{Accreted component: the Bulge-Stellar Halo connection}
\label{chap:accreted}

\begin{figure*}
\includegraphics[scale=0.5]{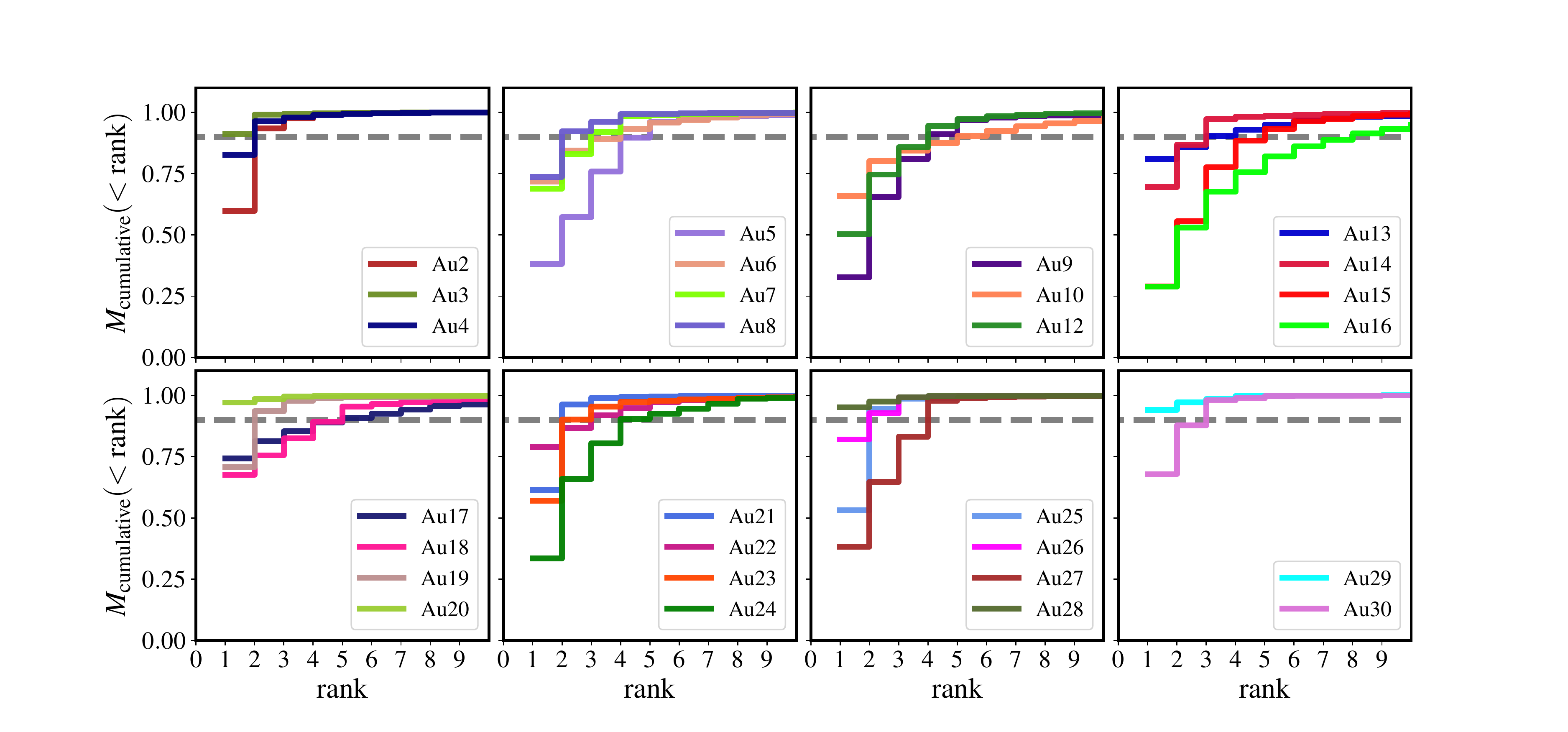}
\caption{Cumulative accreted mass fraction from progenitors ranked by its contribution to the accreted component of the bulge. The higher rank (lower value) corresponds to the progenitor that contributed the most mass to the bulge. 
The gray dashed line represents the $90\%$ of the total accreted mass. Galaxies are presented in small groups in different panels for clarity.}

\label{fig:Cumulative-Acc}
\end{figure*}

In this section, we analyze the accreted component of the Auriga bulges, and its connection to the more extended stellar halo. As previously discussed in Section ~\ref{sec:AccInsituDef}, the accreted component in all our models is subdominant, with a fractional mass that never surpasses the $f_{\rm acc} \lesssim 0.42$, and a median of $f_{\rm acc} = 0.08$.  In Figure ~\ref{fig:Cumulative-Acc} we dissect the total accreted bulge mass into different
satellite contributors. Satellites are ranked according to
their fractional mass contribution to the bulge in decreasing order (i.e. the larger  the contributor, the smaller the rank assigned). 
We find that most of the accreted component ($90\%$) of each simulated bulge is formed from a low number of progenitors, between $1$ for {\rm Au3} and $8$ for {\rm Au16}.  Rather than a continued assembly of mass coming from small satellites, we find the accreted part of the Auriga bulges are mainly built-up from a few major accretion events.

\begin{figure*}
\includegraphics[scale=0.45]{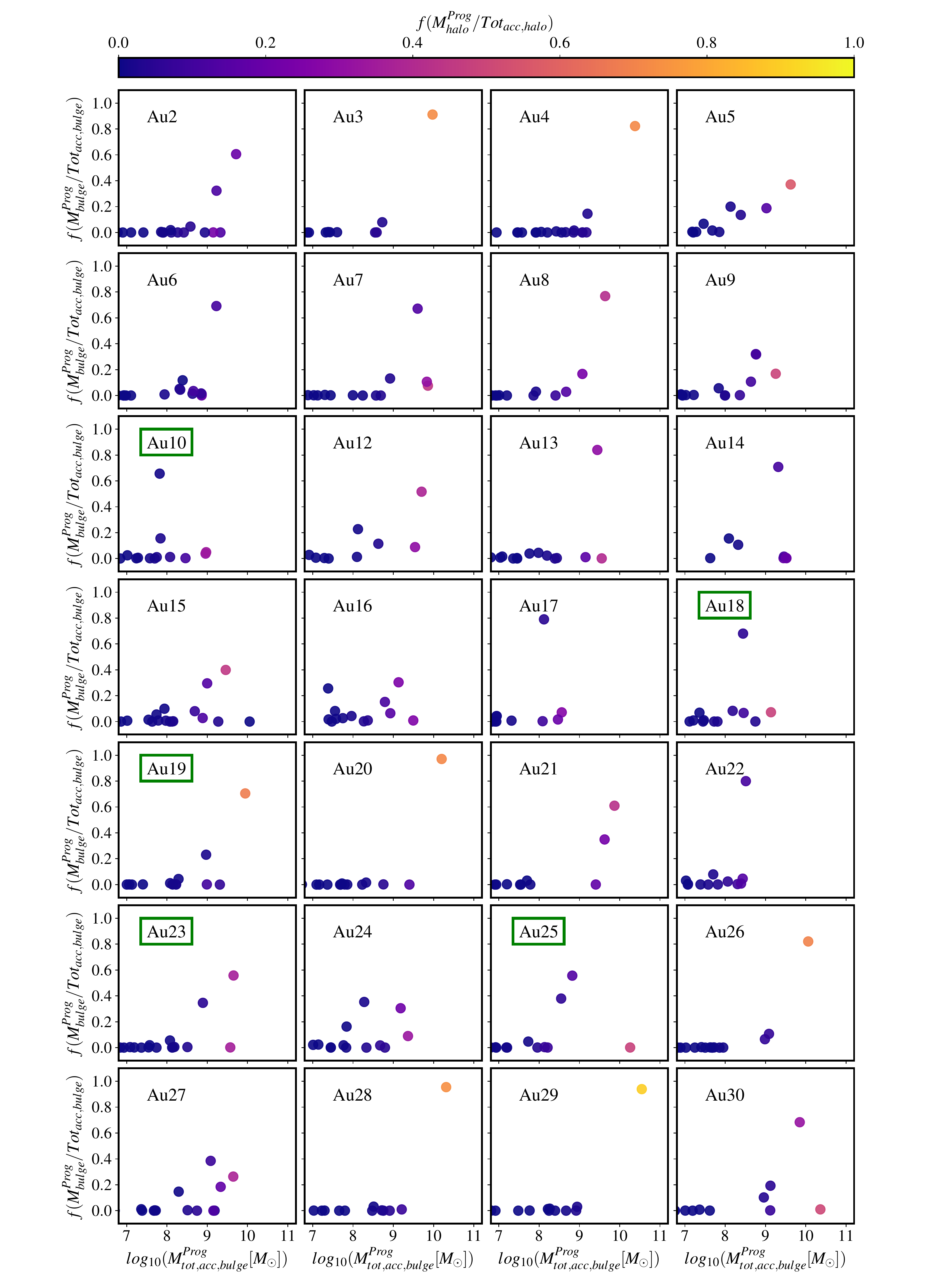}

\caption{ Fraction of mass that each progenitor (with stellar mass higher than $10^7 {\rm M}_\odot$) contributes to the bulge with respect to accreted bulge mass, as a function of the total stellar mass of the progenitor at the time of accretion into the host galaxy. The colouring of the points represents the fraction that each progenitor contributes to the stellar halo with 
respect to the total accreted mass in the halo. Green rectangles in the simulation identifier indicate the galaxies with accreted bulge fraction lower than 1\%.}

\label{fig:accfractions-colors}
\end{figure*}

Now, we take a closer look at the progenitors of the accreted bulge component. In Fig.~\ref{fig:accfractions-colors} we show, for the different progenitors, the contributed mass fraction to the accreted bulge against the total stellar mass associated with each progenitor. We only focus on satellites with stellar masses higher than $10^7 {\rm M}_{\odot}$. In general, more massive progenitors contribute with more mass to the accreted bulge. However, we find several examples (1/3 of the Auriga bulges) in which the most massive progenitors have contributed with either negligible or small fractions to the accreted bulges. Examples are Au10, Au13, AU18, Au25 and Au30. This is typically the case for galaxies with negligible accreted bulge fractions ($f_{\rm acc,bulge} < 0.01$, see Fig.~\ref{fig:Bar-acc-insitu}), for which we show the simulation identifier highlighted with a green rectangle in the corresponding panels. Most of the stellar mass brought in by these massive progenitors was either placed on the stellar halo (M2019) or in a ex-situ disc component \citep{Gomez2017b}. 

During the infall of a satellite galaxy, 
a fraction of their stars are stripped away by tidal forces and scattered 
to  form the stellar halo \citep{SearleZinn1978, BullockJohnston2005, Cooper2010, Helmi1999, Helmi2008}. This places mergers as the main contributors to the formation of stellar haloes. Historically, mergers are considered to play an important role also in the formation of bulges \citep[See][for a review]{BrooksChristensen2015}, so following this scenario one should expect some relation between the growth of the stellar halo and the bulge of a given galaxy. The colour coding in Fig.~\ref{fig:accfractions-colors} indicates the fraction of mass that each progenitor contribute to the total  stellar halo at $z=0$. M2019 already showed that also a few massive progenitors are responsible for the mass assembly of the majority of the stellar halo ($> 90\%$). Here we see that in many cases (see e.g. Au3, Au4, Au19 or Au20), some of the most significant progenitors of the stellar halo are also the most significant progenitors of the bulge. However, as previously stated, several counterexamples can be found.

\begin{figure*}
\includegraphics[scale=0.45]{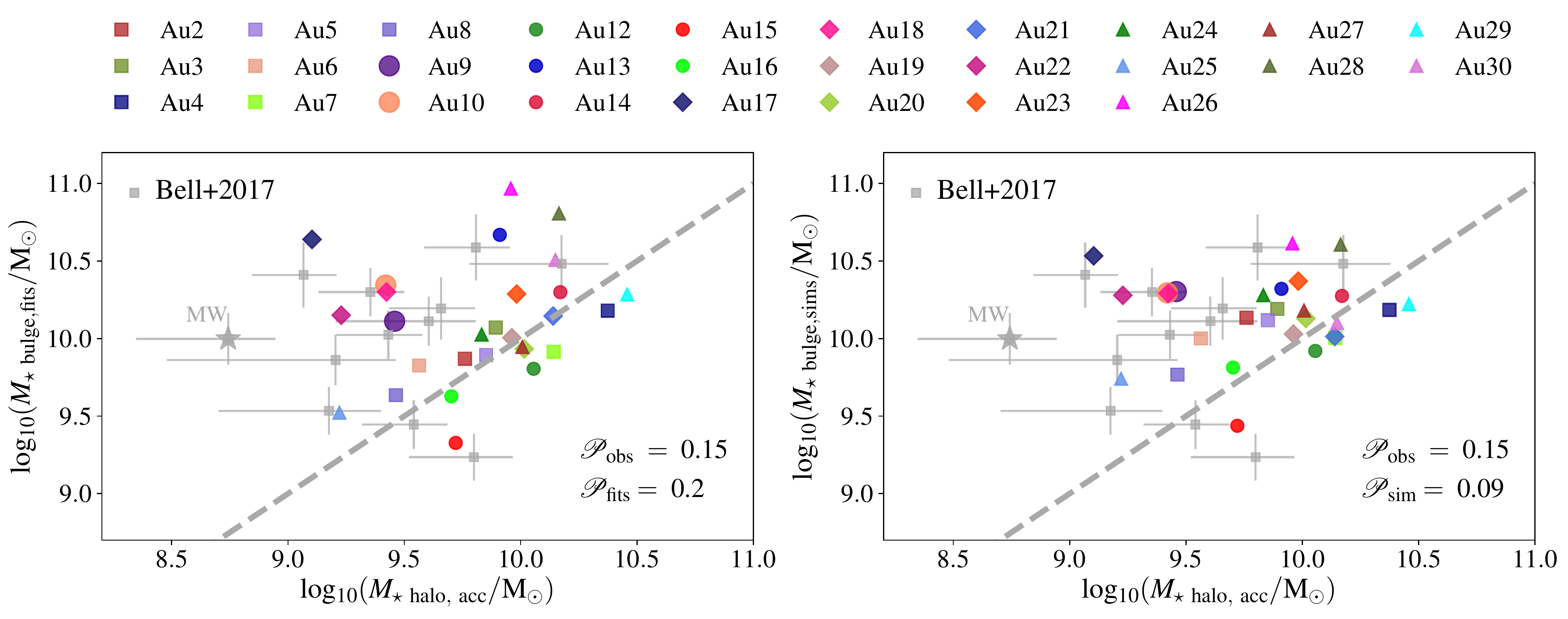}

\caption{{\it Left panel: } Stellar bulge mass as a function of accreted stellar halo mass. Bulge masses are estimated using the ${\rm B/T_{v}}$ ratio derived from surface brightness profiles as in the observations. Each color indicates a galaxy, as labelled at the top. Gray points
with errorbars are observed data from \citet{Bell2017}. The grey dashed line indicates a 1:1 correlation for reference. No correlation in the Auriga bulges can be seen and little to no correlation can be seen in the observed sample. Pearson coefficients for the observed sample and simulated bulges are shown in the bottom right corner. {\rm Au 9}, {\rm Au 10}, {\rm Au17}, {\rm Au22} and {\rm Au 26} show large bulge masses and low mass accreted stellar haloes. {\it Right panel}: Same as in the left panel, but bulge stellar masses are computed as the sum of the mass of stellar particles inside the bulge region defined in Sec.~\ref{sec:bulge-def}.}

\label{fig:Bell-Aurigas}
\end{figure*}

\citet{Bell2017} compares the stellar masses of the bulges and halos of nearby MW-mass galaxies (MW-peers); that is galaxies with stellar masses in the range of $3-12 \times 10^{10} {\rm M}_{\odot}$. The idea of this work was to study the relation between the formation of bulges and the merger history of galaxies.
They found very little correlation
between the accreted mass of the stellar haloes and the total mass of the bulges. 
To make a fair comparison with the observations, we computed the total bulge mass of the Auriga galaxies by multiplying the total stellar mass of the simulations by the bulge-to-total ratio obtained by integrating the fitted functions obtained in the two-component decomposition of surface brightness profiles (see Sec.~\ref{sec:generalprop}) . This is how the total bulge mass was computed for the observed galaxies studied in \citet{Bell2017}. The total stellar mass of each Auriga galaxy in this particular case was computed using the color-dependent mass-to-light ratios from \citet{Bell2003}. The B-V colour and total luminosity in the K-band were evaluated inside the optical radius of our simulations to derive the final total stellar mass. The resulting bulge stellar masses, for our models, computed following this procedure are shown in the left panel of Fig.~\ref{fig:Bell-Aurigas} as a function of the accreted stellar halo mass, obtained from M2019.  In the right panel we show the same relation as in the left panel, but bulge stellar masses were computed as the sum of the stellar particle masses in the bulge as defined in Sec.~\ref{sec:bulge-def}. Our results in both panels reproduce qualitatively those obtained by \citet{Bell2017} for observed galaxies, also shown in Fig.~\ref{fig:Bell-Aurigas} as grey symbols. There is no clear correlation between these two quantities in the Auriga galaxies.  Although this is clear by eye, we computed the Pearson coefficient for the observations and the simulations, which measures the degree of linear correlation taking values between $1$ and $-1$. The Pearson coefficient for the simulations in the right panel is close to 0, indicating a null correlation. When using the luminosity weighted bulge masses in the left panel the Pearson coefficient yields 0.2 and for the observed sample, the value is 0.15, larger, but still low. 
We find a group of galaxies ($\sim 16\%$ of Auriga haloes) with low accreted stellar halo mass and a large bulge mass. This population is also present in the observed sample. 

\begin{figure}
\includegraphics[scale=0.45]{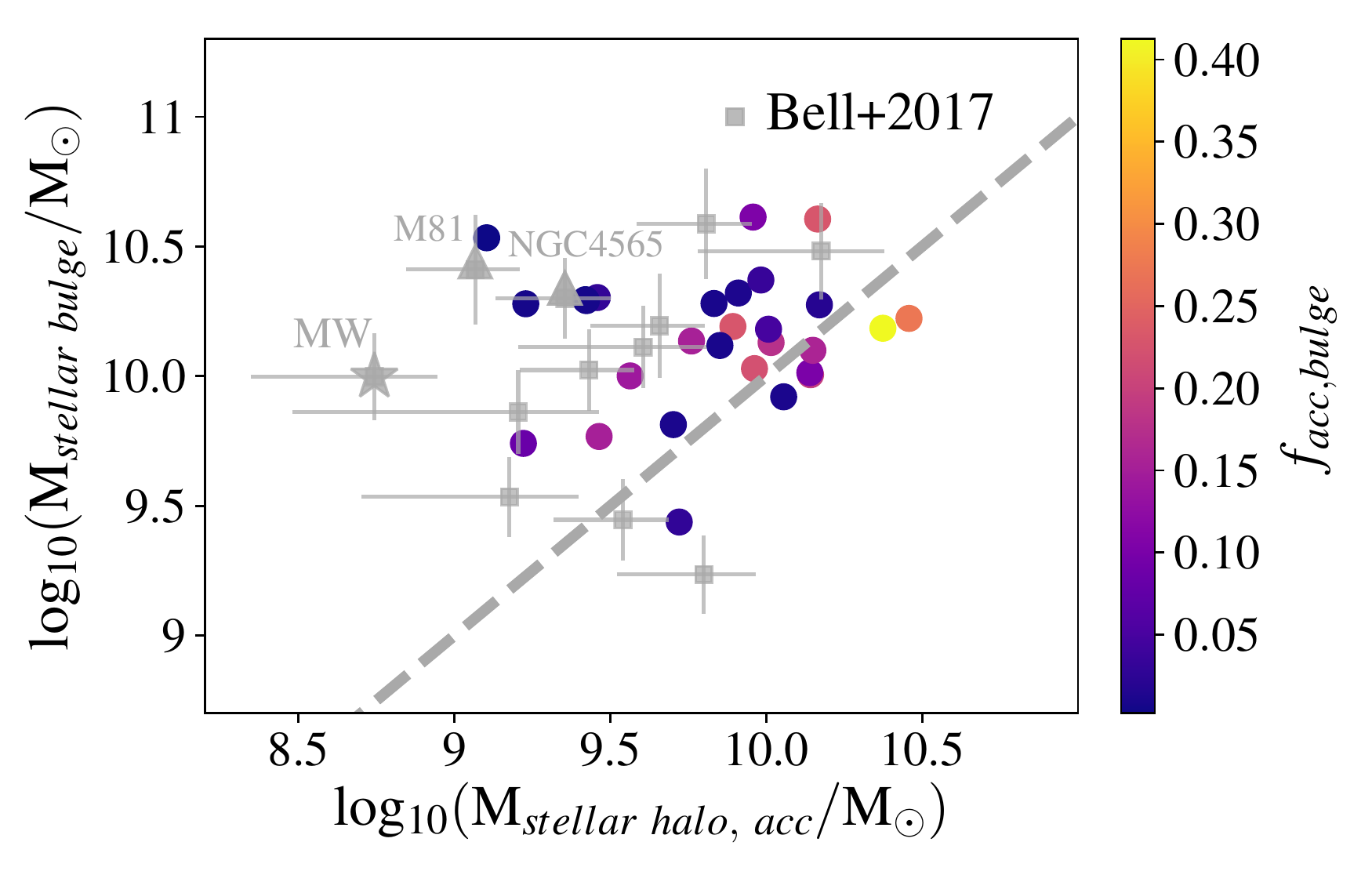}
\includegraphics[scale=0.45]{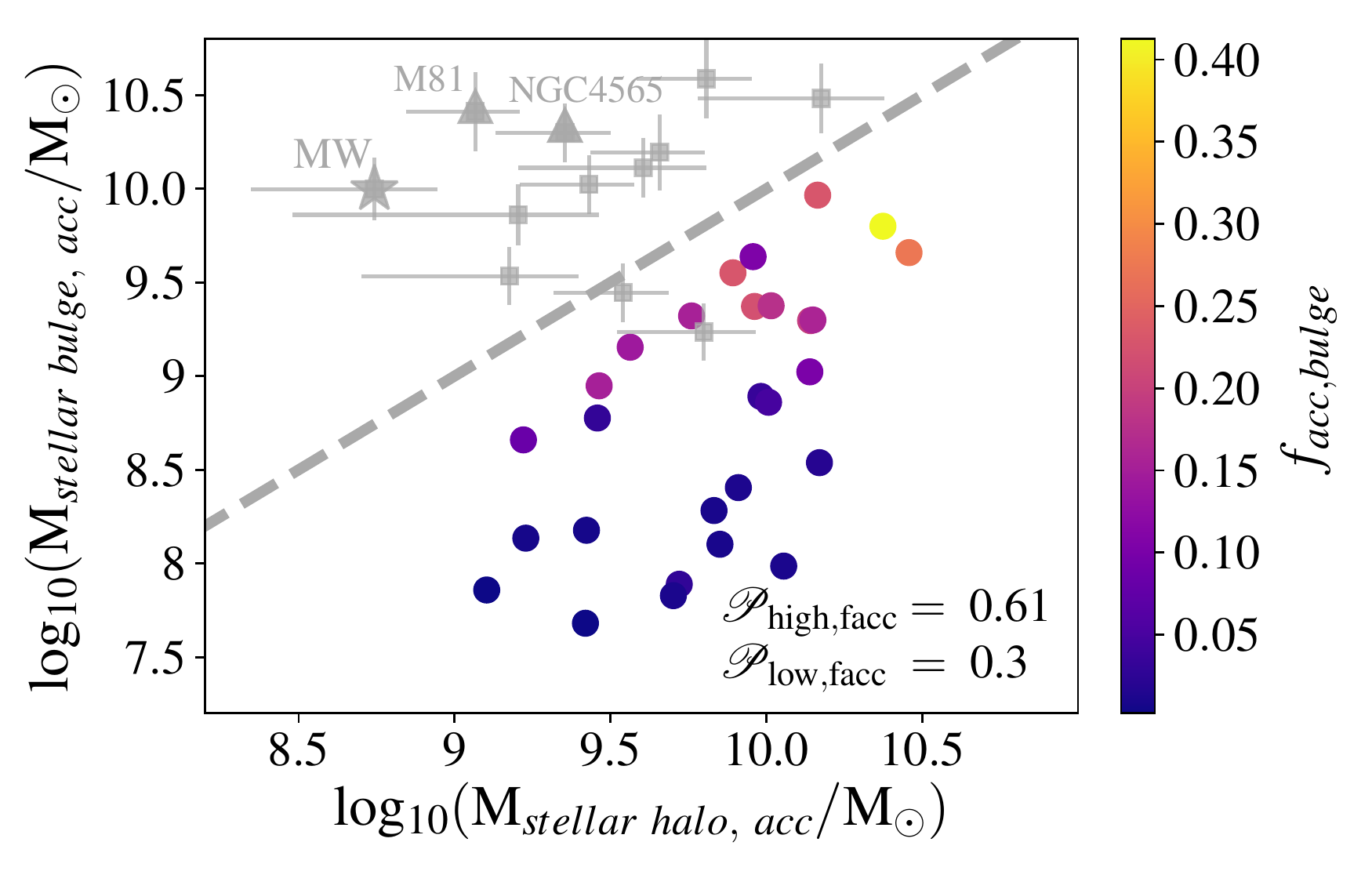}

\caption{Total bulge stellar mass (top) and accreted bulge stellar mass (bottom) as a function of accreted stellar halo mass. Points are coloured
according to the accreted fraction of stars of each bulge. Grey points with errorbars are observed data from \citet{Bell2017} and the grey dashed line indicates a 1:1 correlation for reference. The group of high total bulge mass, mentioned in Fig.~\ref{fig:Bell-Aurigas},
presents marginal accreted bulge fractions. Galaxies with higher accreted bulge fractions show a tight correlation in the bottom panel. Pearson coefficients are shown for galaxies with high and low accreted bulge fraction ($f_{\rm acc,bulge} >0.05$ and $f_{\rm acc,bulge} < 0.05$) respectively.}

\label{fig:Bell-facc}
\end{figure}

To further investigate the origin of the location of the Auriga galaxies in the bulge mass vs. halo mass diagram, in the top panel of Fig. ~\ref{fig:Bell-facc} we have color-coded the symbols according to respective accreted bulge mass fraction. In this case we show the bulge masses computed directly from the simulation as in the right panel of Fig.~\ref{fig:Bell-Aurigas}, since we cannot compute the accreted fractions of the bulge masses derived from the surface brightness profiles.  Interestingly, the five simulations with a high bulge mass and low accreted halo mass (Au9, Au10, Au17, Au18, and Au22 in Fig.~\ref{fig:Bell-Aurigas}) all have negligible accreted bulge fractions. This fact alone, however, does not explain the offset from the broad relation followed by the other galaxies, since there are other galaxies with low
accreted bulge fractions which do follow the weak trend between
total bulge mass and accreted stellar halo mass (as we can see
in Fig.~\ref{fig:Bell-facc}). However, it is important to notice that these are some of the galaxies with the most massive bulges in our set of simulations. 

As previously discussed in Sec.~\ref{sec:insitu}, \citet{Grand2016} recently characterized the disc vertical heating of the Auriga galaxies, associated with non-axisymmetric disc perturbations and merger events.  If we concentrate on the group of bulges with high total bulge mass but negligible accreted bulge fraction, Au17 and Au18 show two of the strongest bars in the whole simulation sample, that are present nearly during the entire evolution of these galaxies (see their Fig. 5). Au9 develops a strong bar from $6~ {\rm Gyrs}$ ago until the present-day. Similar results are found for Au10 and Au22. High total bulge mass and low accreted bulge fraction is also seen in Au13, Au14 and Au23. However, they do follow the broad  $M_{\rm bulge}$ vs $M_{\rm halo}^{\rm acc}$ correlation in the diagram more closely, mainly thanks to their larger $M_{\rm halo}^{\rm acc}$ mass. We find that these galaxies develop relatively strong non-axisymmetric disc features during their evolution. 
In addition, they experienced a relatively massive merger, which contributed to the formation of the in-situ bulge through the triggering of central star formation bursts and tidal perturbation of the pre-existing discs (see Fig.~\ref{fig:insitu-tvsr-1} and Fig.~\ref{fig:insitu-tvsr-2}), but mainly increasing the accreted halo mass, $M_{\rm halo}^{\rm acc}$. This can be seen in Fig~\ref{fig:accfractions-colors}.

Based on our results, one can infer that galaxies with low accreted stellar halo and high total bulge mass (Au9, Au10, Au17, Au18 and Au22) have a low bulge accreted mass fraction, and that non-axisymmetric disc perturbations have played a significant role in the build-up of the total bulge mass. Following this, M81 and NGC 4565, highlighted in Fig.~\ref{fig:Bell-facc} with grey triangles, should have a low fractional contribution of an accreted bulge component with respect to the in-situ bulge component and that, probably, they have developed strong non-axisymmetric features, such as bars or transient spiral arms, sometime during their evolution. NGC 4565 was studied recently by \citet{KormendyBender2018} and indeed fit this scenario showing lack of a classical bulge component and the presence of an X-shape central region attributed to a bar. M81, for its part, does not show a visible bar, but it is known for its strong spiral pattern \citep{Kendall2008}. Considering our findings, a possible scenario for the apparent excessive mass of M81's bulge in relation with its stellar halo mass is that an ancient bar might have developed sometime in this galaxy.

In the bottom panel of Fig.~\ref{fig:Bell-facc} we show the accreted bulge mass as a function of the accreted stellar halo mass, color-coded by the accreted bulge fraction. We can see that the accreted bulge mass correlates with the accreted halo mass. Bulges with high accreted fraction show a tight correlation, close to a 1:1 correlation, whilst bulges with low and negligible accreted fractions show a larger scatter. In order to quantify the difference, we computed the pearson coefficients obtained for galaxies with accreted bulge fractions $f_{\rm acc} > 0.05$ and for those with lower accreted bulge fractions separately. The high accreted bulge fraction group shows a pearson coefficient of 0.76 and the group with low accreted bulge fraction has a pearson coeficcient of 0.42, as indicated in the bottom right corner of the bottom panel.

\section{Discussion}
\label{sec:discussion}

Historically, simulations of galaxy formation within a $\Lambda$-CDM framework suffered the overcooling problem \citep{Balogh2001}. Simulations produced highly concentrated galaxies with bulges that exceeded the observed sizes and bulge-to-total ratios \citep{ChristensenBrooks2014}. 
With the implementation of plausible sub-grid physics in hydrodynamical simulations  that take into account energetic feedback processes, the excess of concentrated star formation
was suppressed \citep{Springel2003, CeverinoKlypin2009}, although, sometimes, at the expense of 
destroying the morphology of the galactic discs \citep{Roskar2014}. 
In the case of the Auriga
galaxies, the stellar feedback produces a moderate thickening of the discs but preserves properties that reproduce several observational trends (G2017). With regard to the bulges, we found that the Auriga galaxies show a systematic excess in the effective radii related to an excess in the total mass of bulges. Because of this excess, 23\% of Auriga bulges have surface brightness inside effective radius and total Magnitude with no counterpart in nature. This excess could be related to the implementation of the stellar feedback in the Auriga simulations, which has been shown to produce a milder effect than expected in the quenching of dwarf galaxies  \citep{Simpson2018}. 

One of the main results of this paper is the predominant in-situ formation of the Auriga bulges.
In a recent work, \citet{Gargiulo2017} studied the stellar populations of galactic bulges in MW-like galaxies using
a semi-analytic model of galaxy formation \citep{Gargiulo2015, Cora2018}, coupled to cosmological N-body simulations, and found that on average approximately~$16\%$ of the bulge mass has an accreted origin.
Other cosmological simulations show similar results. \citet{Tissera2018} show the fraction of accreted mass as a function of galactocentric distances in their Fig. 2, and  even though the inner spheroid  includes some transitional region to the
stellar halo, we can see that the accreted fractions inside 4 {\rm kpc} are also between $0.02$ and $0.15$.  \citet{Guedes2013} reports an accreted fraction of only $4\%$ in the pseudo-bulge of a MW-like galaxy simulation. \citet{Buck2018b} found a contribution of $2.3\%$ of accreted stars to the bulge components. Ours and previous results  suggest that bulges of simulated disc galaxies formed in MW-sized DM haloes in a cosmological context commonly lack a strong accreted component.

When we look exclusively at the accreted stellar 
particles that are part of bulges at $z=0$ we found that a low number of satellite galaxies (the most massive progenitors in the history of accretion in general) are responsible for the build-up of the accreted bulge component of galaxies (see Fig.~\ref{fig:Cumulative-Acc} and Fig.~\ref{fig:accfractions-colors}). This result suggests that many mergers with very small satellites are not centrally responsible for the growth of galactic bulges. A similar result was found for the stellar haloes of the Auriga simulations by M2019. As also noted by M2019, this is in line, for example, with the formation scenario of the stellar halo of M31, for which a single satellite is found to account for the build-up of most of its stellar halo \citep[see also][]{DeSouza2018, DeSouza2018b}. Following our findings, this single satellite could also be responsible for most of the accreted component of its bulge.
Although one might at first think that a halo or accreted component of a bulge built from one or a few progenitors would have relatively homogeneous populations, in fact, the halos built from few massive progenitors have complex star formation histories and gradients inherited from the massive satellite that it accreted \citep[][]{DeSouza2018, DeSouza2018b}.

Mergers are a natural outcome of the $\Lambda$-CDM scenario, in which DM haloes, where galaxies are formed, growth hierarchically from smaller structures. Theoretical predictions of merger rates are in close agreement with estimates of merger rates using pair fractions. \citet{Mundy2017} used this method to estimate
observational merger rates and compared with results using the ILLUSTRIS simulation \citep{Rodriguez-Gomez2017}, and merger rates derived from light cones derived from a semi-analytic model \citep{Henriques2015}, and found a good agreement. The influence of
mergers in the formation of bulges is found to be central in many simulations.  Major mergers are common at high redshift, but after $z\sim1$ become rare for galaxies with masses in
the range of the MW-mass.  \citet{GarrisonKimmel2018} show using the LATTE simulations \citep{Wetzel2016}, that most of the stars in their simulations are born with disc-like kinematics and that gas rich mergers are one of the main drivers of bulge formation at high redshift.

It is often claimed that the hierarchical paradigm is in tension with the lack of classical bulges in bright disc galaxies like the MW \citep[see, for example][]{Shen2010, Kormendy2010, Kormendy2015}.
It is important to highlight the fact that most large disc galaxies in the local Universe have pseudo-bulges is not at odds with galaxy formation in a hierarchical context, as can be seen in the results of our study.
Results presented in this work show that mergers funnel gas into the central parts 
of the galaxy forming bulges that do not necessarily show a dominant
dispersion dominated component that resembles a mini-elliptical galaxy at present, as indicated by the properties of the surface brightness profile at z=0, such as the S\'ersic index (see Table~\ref{table:bulgeprop}), its shape (see Fig. ~\ref{fig:coaboadiagram}) and the degree of ordered rotation (see Fig.~\ref{fig:vosigma-elip}). Even though most of the Auriga simulations present rich merger histories and some of the simulated bulges have large accreted fractions (for example Au4), we found, applying criteria based on those used in observations that the majority of bulges in our sample can be classified as pseudo-bulges.

Bulges formed by mergers can change their shape during their evolution.
It is possible that the formation of a bar on top of a small classical bulge formed at high redshift can 
dynamically affect the non-rotating component, making it indistinguishable from a pseudo-bulge at $z=0$, as is demonstrated by simulations \citep{Saha2012, Saha2015} and as can be seen clearly in the case of Au2 in Fig.~\ref{fig:AccIns-DensShape1}. 
Moreover, mergers or interactions with satellite galaxies can play an important role in the formation of pseudo-bulges, by instigating the formation of bars in globally unstable systems \citep{Byrd1986, Noguchi1987, Romano-Diaz2008}. We see this occurring in some Auriga galaxies, see e.g. the case of the formation of the bulge in Au18 which was discussed in Sec.~\ref{sec:insitu}. In that case, a strong bar is developed after a major merger.
Because of this, we stress that studying the formation of the innermost regions of galaxies, where bulges arise, without a hydrodynamical cosmological framework could lead to oversimplified scenarios of the bulge formation in the cases when the accreted component of the galaxy is significant.

As stated before in Sec.~\ref{sec:methodology}, the question of environment could play a role here. Although the accretion histories were not explicitly constrained, and a good diversity of accretion histories is available, the Auriga simulations were run adopting an isolation criterion for the host DM halos. MW-analogs in denser environments might suffer later and more numerous accretion events on average and, therefore, contain central regions with a higher degree of random motions. Bars can be destroyed by these interactions, limiting important channels of mass formation via secular processes described in Sec.~\ref{sec:insitu}.

\section{Summary and Conclussions}

We have analysed the properties and the origin of galactic bulges in 30 cosmological
magneto-hydrodynamical simulations of disc galaxies in 
MW-sized DM haloes. The Auriga simulations have a baryonic resolution of $\sim 5 \times 10^4 {\rm M}_{\odot}$, and represent one of the largest samples of cosmological collisional simulations at this level of resolution to date. 
We found a great diversity of
bulges in a narrow range of DM halo masses.  This diversity is similar to that observed and should be regarded as a good asset to learn about the formation history of the MW and MW-analogs, such as M31, NGC 4565 and NGC 5746 \citep{KormendyBender2018, Bell2017}. We list here our main results. 

\begin{itemize}

    \item Galactic bulges in the Auriga simulations have considerable diversity in general properties such as morphology, surface brightness, and shape. However, compared with the sample of disc galaxies in the local Universe of \citet{FisherDrory2008}, 
    the Auriga bulges occupy a rather narrow range in size in  the Kormendy diagram. They show a systematic excess in effective radii compared with the observed sample at high surface brightness.

    \item All bulges in the Auriga galaxies have properties that would define them photometrically as pseudo-bulges. S\'ersic indices derived from surface brightness profiles are found to be all $n < 2$ and their behaviour in the $M_v - \mu_e$ diagrams is closer to that observed for pseudo-bulges. Although, $23\%$ of bulges have higher surface brightness than any observed bulge and are more luminous than observed pseudo-bulges.
    
    \item Auriga bulges occupy two marked loci in a
    diagram that relates the axis ratios of their mass distributions. One group shows marked prolate intrinsic shapes due to the presence of a bar, while the other has close to spherical mass distributions. We found that the more nearly to spherical group has lower ${\rm B/T_{sim}}$ ratios and higher S\'ersic indices. Prolate bulges develop in barred galaxies and are in general more massive and have lower S\'ersic indices. 
    
    \item Auriga bulges show a high degree of ordered rotation and low ellipticities in the $V/\sigma-\epsilon$ diagram, well above the region occupied by observed classical bulges. 
    
    \item Auriga bulges are formed predominantly in-situ (see Fig.~\ref{fig:Bar-acc-insitu}). The largest accreted bulge fraction reaches $f_{\rm acc} = 0.42$, and only $21\%$ of bulges have accreted fractions higher than $f_{\rm acc} = 0.2$. $21\%$ of bulges have negligible accreted fractions (less than $f_{\rm acc} = 0.01$). 
    
     \item The spatial distributions of the accreted and in-situ bulge components show different behaviours. $50\%$ of the Auriga bulges possess a less concentrated accreted component compared to the in-situ component, whereas $11\%$ of bulges show that these components have a similar normalised spherical density distribution. Interestingly, $14\%$ of bulges show accreted components that follow the shape of the bar seen in the corresponding in-situ components. The remaining bulges have too low or negligible accreted fractions to make a meaningful comparison.
    
    \item Analysis of the in-situ component of bulges show that bulge stellar particles form predominantly in the central regions of the galaxy, although a low fraction of the sample show approximately equal parts formed outside and inside $2 r_{\rm eff}$ at $z=0$. Some of them that happen to develop a strong bar during the evolution form more stars outside the bulge radius than inside. This underscores the significance of secular evolution to the growth of bulges in these galaxies.  Mergers are also responsible to the in-situ star formation inside the bulge. In situ star formation histories of the Auriga bulges show peaks when a single or several minor mergers occur between snapshots. The strength of these peaks correlates broadly with the mass of the satellites.
    The star formation history of the in-situ
    bulge component shows greater diversity, due to the contribution of different mechanisms to in-situ bulge growth and the inherent diversity of merger histories. 
    
     \item The accreted components of bulges in the Auriga simulations are dominated by a  single accretion in 80\% of the cases. Most of the remainder of the mass originates from the next few most massive accretions, which happen to be the more massive satellites that were accreted by the galaxies, in most cases. In addition, the same satellites typically contributed more to the build-up of the corresponding stellar halo. Accretion of stars by a large number of mergers with very small satellites is therefore not a favoured mechanism of the accreted bulge growth in MW-sized galaxies in our models. However, as mentioned before, minor mergers play an important role in the in-situ star formation inside the bulge.
    
    \item Total bulge mass and accreted stellar halo mass show little correlation as noted in observations. Those galaxies with a low accreted bulge fraction show no correlation between stellar halo mass and total bulge mass, but a group of them that develop strong bars during its evolution, occupy a well defined region in the bulge mass vs. accreted stellar halo mass diagram. This result can help to constrain the origin of bulge stars in observed galaxies according to the position in this diagram. Accreted bulge mass and accreted stellar halo mass are correlated. Galaxies with higher accreted bulge fraction show lower dispersion than galaxies with low accreted bulge fraction in this relation. 
    
    This is the first paper in a series studying the
formation of bulges in MW-sized DM haloes.  
    An even larger sample of high resolution simulations
spanning a broader range in DM halo masses is necessary to understand the transition in the formation of
pseudo-bulges and classical bulges in disc galaxies.

\end{itemize}

\appendix 
\section{Two-component decomposition of surface brightness profiles}
\label{sec:appendix}

\begin{figure*}
\includegraphics[scale=0.5]{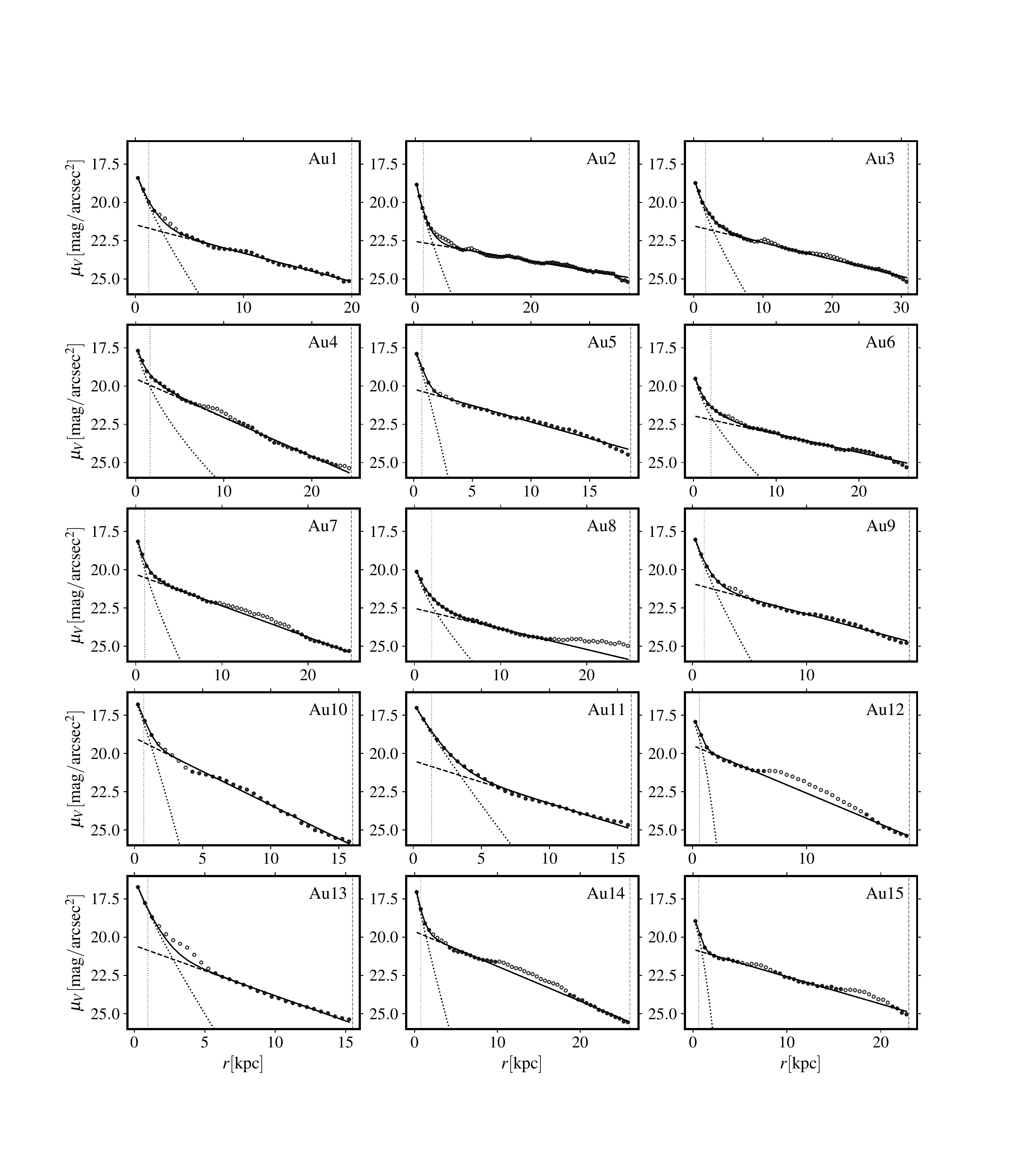}

\caption{Bulge-disc decompositions of all the Auriga simulations from surface brightness fits that simultaneously fits a S\'ersic profile and an exponential profile. Features in the surface brightness profiles due to bars, spiral structure, and rings, that deviate from the smooth nature of the model described by the sum of a S\'ersic model (shown as dotted line) and an exponential profile (shown as dashed line) are excluded from the fit. We show with empty circles the values excluded from the fitting procedure. The vertical dotted line indicates the effective radius of the bulge component. The dashed vertical line indicates the optical radius of the simulation.  }

\label{fig:sbprof-1}
\end{figure*}

Here we show the two-component decomposition of surface brightness profiles of the Auriga simulations. Surface brightness profiles were computed with the simulations as seen face-on, in concentric  annuli of 500 {\rm pc} wide, centered in the coordinate origin, defined as the most bound dark matter particle of each simulation.
We use a non-linear least-square method to fit the sum of an exponential profile and a S\'ersic model \citep{Sersic1968} which in terms of intensity reads:

\begin{equation}
I(r)=I_{\rm e}\exp\big\{ - b_{\rm n} \big[(r/r_{\rm eff})^{1/n} -1  \big] \big\} + I_{0}\exp\left[- (r/R_{\rm scale}) \right ]\, ,
\end{equation}
\label{eq:serexp}

\noindent where $r_{\rm eff}$ is the effective radius that
encloses half of the total light of the S\'ersic model, $I_{\rm e}$ is the intensity of the bulge component at $r_{\rm eff}$, $n$ is the S\'ersic index, $I_{\rm 0}$ is the central intensity of the disc component, and $R_{\rm scale}$ is the disc scale radius. $b_{\rm n}$ is such that $\Gamma(2n) = 2\gamma (2n, b_{\rm n})$, where $\Gamma$ is the complete gamma function. The intensity is later converted into surface brightness to perform the fit. The fit of the surface brightness profile is limited to the optical radius as in G2017, who performed a similar fit to the mass density profile, following \citet[][]{Marinacci2014}. 
The fitting procedure was performed following \citet{FisherDrory2008}. In their work, surface brightness excesses due to features like bars, lenses, rings, and bright spiral structures are not taken into account in the fitting procedure. The central nuclei is also excluded from their fits. In the following, we exclude from our analysis the points in the surface brightness profile where the bar and other features, like spiral arms and rings are conspicuous. These surface brightness excesses are considered deviations from the smooth surface brightness profile assumed by our model (Eq.~\ref{eq:serexp}). However, we do not exclude the central nuclei because the resolution of the simulations does not allow us to separate this component from the underlying peak of surface brightness. Thus, the total magnitudes of the Aurga bulges may be overestimated compared with the observed ones.  
In Fig.~\ref{fig:sbprof-1} we show the surface brightness profiles of the Auriga simulations. The data used in the fit are shown with black filled circles and those excluded from the analysis with black empty circles. The fitted function is shown with a black line and the corresponding S\'ersic and exponential profiles are shown with dotted and dashed lines, respectively. 

\begin{figure*}
\includegraphics[scale=0.5]{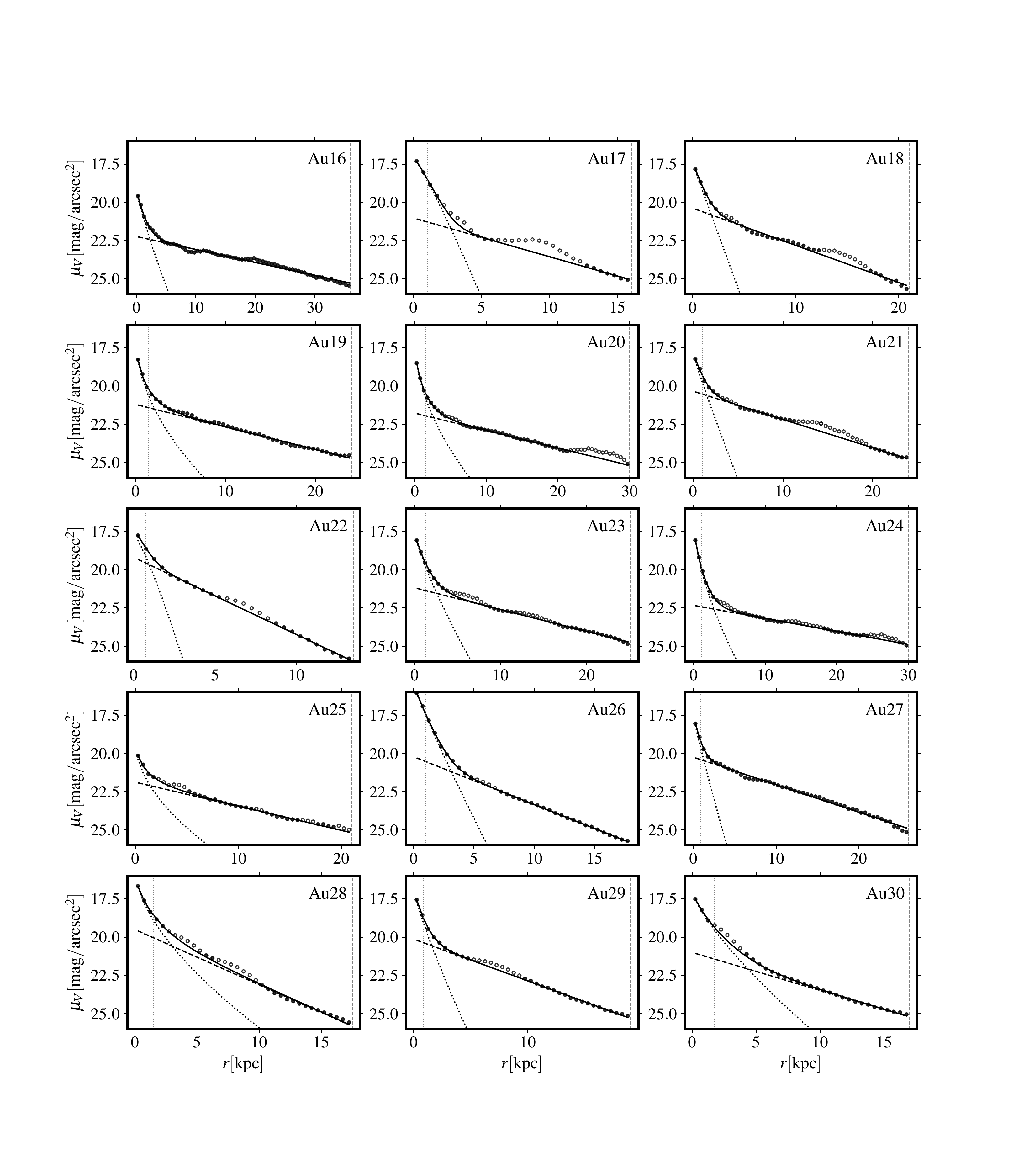}

\caption{Same as in Fig.~\ref{fig:sbprof-1}, but for Au16-Au30}

\label{fig:sbprof-2}
\end{figure*}

\begin{figure*}
\includegraphics[scale=0.5]{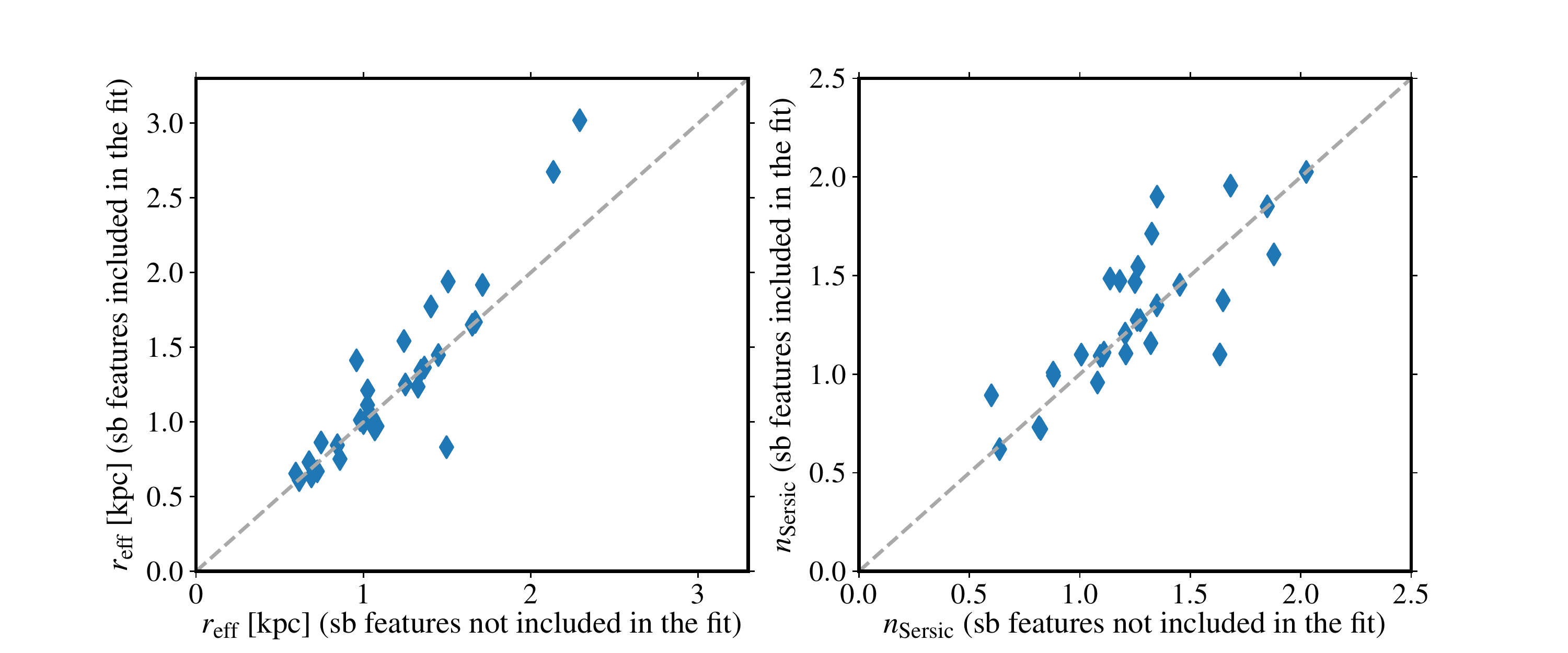}

\caption{{\it left}: Comparison of the effective radius of Auriga simulations obtained when surface brightness(sb) features are excluded from the fit and when they are not excluded from the fit. The dashed line indicate a 1:1 correlation. {\it right}: Same as in the left panel, but comparing the S\'ersic index.}

\label{fig:rnboth}
\end{figure*}

In order to study the robustness of this method, we compare in Fig.~\ref{fig:rnboth}  the S\'ersic index and $r_{\rm eff}$ obtained when  excluding points in the fitting procedure and those obtained using all the available points. We found that effective radii present small differences when using different fitting procedures, showing lower values, in general, when deviations from the smooth model are included in the fitting of the surface brightness profiles.  S\'ersic 
indexes also present differences when adopting different methods to fit the surface brightness profiles, but none of them is larger than n=2 and none of the conclusions drawn in this work are affected by the fitting method used. 

\section*{Acknowledgements}
We wish to thank David Campbell and Adrian Jenkins for generating the initial conditions and selecting the sample of the Auriga galaxies.
I.G. acknowledges financial support from CONICYT Programa de Astronom\'ia, Fondo ALMA-CONICYT 2017 31170048. AM acknowledges support from CONICYT FONDECYT Regular grant 1181797.
FAG acknowledges support from CONICYT FONDECYT Regular grant 1181264. IG, AM and FAG acknowledge funding from the Max Planck Society through a “Partner Group” grant. FM acknowledges support through the program “Rita Levi Montalcini” of the italian MIUR.
This work used the DiRAC Data Centric system at Durham University, operated by the Institute for Computational Cosmology
on behalf of the STFC DiRAC HPC Facility (www.dirac.ac.uk).
This equipment was funded by BIS National E-infrastructure capital
grant ST/K00042X/1, STFC capital grant ST/H008519/1, and
STFC DiRAC Operations grant ST/K003267/1 and Durham University.
DiRAC is part of the National E-Infrastructure.
\bibliographystyle{mnras}
\bibliography{references} 

\label{lastpage} 
\end{document}